\documentclass[12pt]{article}
\usepackage{authblk}
\usepackage{natbib}
\usepackage{amsmath, amsfonts,mathrsfs}
\usepackage{color}
\usepackage{rotating}
\usepackage{JASA_manu1}
\usepackage{lscape}
\usepackage{graphicx,multirow}
\usepackage{bm,url}
\usepackage{booktabs}
\usepackage{caption}
\usepackage{subfig}

\newtheorem{lemma}{\noindent\mbox{Lemma}}
\newtheorem{theorem}{\noindent\mbox{Theorem}}
\newtheorem{corollary}{\noindent\mbox{Corollary}}
\newtheorem{proposition}{\noindent\mbox{Proposition}}
\newtheorem{condition}{\noindent\mbox{M}}
\renewcommand{\theequation}{\thesection.\arabic{equation}}

\DeclareMathOperator*{\argmax}{arg\,max}
\def\be{\begin{equation}}
\def\ee{\end{equation}}
\def\bea{\begin{eqnarray}}
\def\eea{\end{eqnarray}}
\def\bd{\begin{displaymath}}
\def\ed{\end{displaymath}}
\def\bda{\begin{eqnarray*}}
\def\eda{\end{eqnarray*}}
\def\bsm{\begin{small}}
\def\esm{\end{small}}

\def\nn{\nonumber}

\newcommand{\E}{\rm E}
\newcommand{\V}{\rm Var}

\def\ha1{\hat \beta_1}

\def\bb0{\delta_\beta}

\def\bsc{\begin{scriptsize}}
\def\esc{\end{scriptsize}}

\begin{document}

\title{A Neighborhood-Assisted Hotelling's $T^2$ Test for High-Dimensional Means}
\author[1]{Jun Li}
\author[2]{Yumou Qiu}
\author[1]{Lingjun Li}
\affil[1]{Department of Mathematical Sciences, Kent State University, Kent, OH 44242}
\affil[2]{Department of Statistics, University of Nebraska-Lincoln, Lincoln, NE 68583}

\renewcommand\Affilfont{\itshape\normalsize}
\date{}
\maketitle

\begin{center}
\textbf{Abstract}
\end{center}
Many tests have been proposed to remedy
the classical Hotelling's $T^2$ test in the ``large $p$, small $n$" paradigm, but the existence of an optimal sum-of-squares type test has not been explored.  This paper shows that under %a non-local alternative condition, 
certain conditions, the population Hotelling's $T^2$ test with the known $\Sigma^{-1}$ attains the best power among all the $L_2$-norm based tests with the data transformation by $\Sigma^{\eta}$ for $\eta \in (-\infty, \infty)$. To extend the result to the case of unknown $\Sigma^{-1}$,  
we propose a Neighborhood-Assisted Hotelling's $T^2$ statistic obtained by replacing the inverse of sample covariance matrix in the classical Hotelling's $T^2$ statistic with a regularized covariance estimator. Utilizing a regression model, we establish its asymptotic normality under mild conditions. We show that the proposed test is able to match the performance of the population Hotelling's $T^2$ test under certain conditions, and thus possesses certain optimality. 
Moreover, it can adaptively attain the best power by empirically choosing a neighborhood size to maximize its signal-to-noise ratio.
%Moreover, the test has the ability to attain its best power possible by adjusting a neighborhood size to unknown structures of population mean and covariance matrix. 
%it provides a framework to unify the performance of two oracle tests: one is shown to be optimal in power under a non-local alternative condition and the other can possess better power under a local alternative condition. 
%An optimal neighborhood size selection procedure is proposed to maximize the power of the Neighborhood-Assisted $T^2$ test via maximizing the signal-to-noise ratio. 
Simulation experiments and case studies are given to demonstrate the empirical performance of the proposed test.

\bigskip

%\vspace*{.1in}


%\newpage
\setcounter{section}{1} \setcounter{equation}{0}
\section*{\large 1. \bf Introduction}

With the explosive development of high-throughput technologies, high-dimensional data characterized by simultaneous measurement of a large number of variables become feasible. Examples include microarray, next-generation sequencing (RNA-Seq), genome-wide association (GWA) and genomic selection (GS) studies, where tens of thousands of genes are measured from any single experimental subject. Others can be seen in financial risk management and marketing research, where the number of the assets in a portfolio or the number of items a consumer purchases are typically very large. Different from traditional small or median-scale data, high-dimensional data involve a considerably large number of variables but relatively small sample size, which makes it very challenging to analyze. 

This article considers testing whether a high-dimensional mean vector is zero, which is one of the conventional statistical problems that face the high-dimensional challenge. Let 
\begin{eqnarray}
X_i=\mu+\varepsilon_i, \qquad \varepsilon_{i}\stackrel{i.i.d.}{\sim}\mbox{N}(0, \Sigma)\, \, \, \mbox{for} \,\,i=1,\cdots, n, \label{model}
\end{eqnarray}
where $\mu = (\mu_{1}, \cdots, \mu_{p})^{\prime}$ is a $p$-dimensional population mean vector, and $\Sigma = (\sigma_{j_1j_2})_{p \times p}$ is a $p \times p$ covariance matrix. Note that the Gaussian assumption in (\ref{model}) is not essential and the results developed in this paper can be extended to non-Gaussian distributions. Here, we are interested in testing the hypotheses
\be
H_{0} : \mu= 0  \quad  \mbox{vs} \quad H_{a} : \mu
\ne 0. \label{eq:hypo1}
\ee

The testing problem is motivated by recent development in genomic studies. It has been believed that each gene does not function individually, but instead cooperates with others sharing common biological functions to achieve certain biological processes (see Barry, Nobel and Wright 2005, Efron and Tibshirani 2007 and Newton et al. 2007). It is therefore more relevant to identify significant gene-sets/pathways rather than a single gene with respect to certain treatments. If we let $\mu$ be the difference of population gene expression levels before and after a certain treatment, testing hypotheses (\ref{eq:hypo1}) can be thought as identifying differentially expressed gene sets.

Let $\bar{X}=\frac{1}{n}\sum_{i=1}^{n}X_{i}$ be the sample mean and
$S_n=\frac{1}{n} \sum_{i=1}^{n} (X_{i}-\bar{X})(X_{i}-\bar{X})^{\prime}$
be the sample covariance matrix.
When $p$ is fixed and $p \le n-1$, Hotelling's $T^2$ statistic
\be
T^2=n\bar{X}^{\prime}S_n^{-1}\bar{X} \label{eq:HT}
\ee
has been used for testing the hypotheses (\ref{eq:hypo1}).
With Gaussian data, $(n - p)T^2/\{(n-1)p\}$ follows a central F-distribution with degrees of freedom $p$ and $n-p$ under $H_{0}$ of (\ref{eq:hypo1}). Furthermore, the Hotelling's $T^2$ test is uniformly most powerful among all tests that are invariant with respect to the transformation $C X_i$ for a nonsingular matrix $C$ (Anderson, 2003). Despite its optimality, the Hotelling's $T^2$ test encounters an unsatisfied performance when $p>n-1$ because of the singularity of $S_n$. Even when $p<n-1$ and $p$ is close to $n-1$, the Hotelling's test loses its power as revealed by Bai and Saranadasa (1996). 

Many proposals have been considered to correct Hotelling's $T^2$ test in high dimensional setting. Some were constructed by excluding or stabilizing $S_n^{-1}$ to avoid its adverse effect. Examples include Bai and Saranadasa (1996), Srivastava and Du (2008), and Chen and Qin (2010). More can be seen in Wang, Peng and Li (2015), Chen et al. (2011), and Li et al. (2016). %that is utilized to incorporate data dependence, because it is not a consistent estimator under high dimensionality. However, simply ignoring dependence in the test statistics will lose a large part of the mutual information contained in the data. 
A drawback of the above tests is that they discard the dependence information among variables, which may suffer power loss. 
On the contrary, thresholding tests and maximum type tests that incorporate covariance information were proposed for sparse signal detection. Examples include Hall and Jin (2010), Cai, Liu and Xia, (2014) and Chen, Li and Zhong (2015). %Despite their established optimality, 
However, those tests are not as powerful as the sum-of-square tests when signals are weak and dense. Moreover, the implementation of the above testing procedures requires strict sparsity structure assumptions on the covariance or precision matrix, %such as the polynomial off-diagonal decay in Hall and Jin (2010). In real applications, these assumptions 
which may not be satisfied in real applications. Different from those approaches, the tests in Thulin (2014), and Srivastava, Li and Ruppert (2016), were proposed to apply the classical Hotelling's $T^2$ statistic to low-dimensional space obtained by data projection. Without prior information on the best low-dimensional space, the tests heavily rely on multiple random projections and thus can be computationally expensive.  

%have been proposed to detect sparse and weak signals by exploiting advantageous effect of  dependence (Hall and Jin, 2010; Cai, Liu and Xia, 2014; and Chen, Li and Zhong, 2015). By showing that the signal strength can be enhanced by transforming the original data through the precision matrix, the proposed tests possess higher power than other $L^2$-norm tests without utilizing data dependence.

Despite many efforts for (\ref{eq:hypo1}) in high dimensional setting, the question of whether an optimal sum-of-squares type test exists has not been explored. To make it a well-defined problem, we first define a class of sum-of-squares type statistics by (\ref{Oracle}) in Section 2. This is analogous to the classical Hotelling's $T^2$ statistic which is restricted to a class of statistics that are invariant with respect to the transformation $C X_i$ for a nonsingular matrix $C$. Another reason we consider (\ref{Oracle}) is to study the impact of data dependence via the covariance matrix $\Sigma$ on the $L_2$-norm statistics. Within the class (\ref{Oracle}), we show that under certain conditions, the population Hotelling's $T^2$ test with the known $\Sigma^{-1}$ is optimal. 

To extend the result to the case of unknown $\Sigma^{-1}$, we propose a new test statistic obtained by replacing $S_n^{-1}$ in the classical Hotelling's $T^2$ statistic with a regularized covariance estimator through banding the Cholesky factor. We show that the test statistic can be interpreted by a linear regression model in which each component of the random vector $X_i$ in (\ref{model}) is regressed to its nearest predecessors. Utilizing the regression interpretation, we establish the asymptotic normality of the proposed test statistic under some mild conditions. The advantages of the proposed test are multifold. First, it has the ability to attain its best power possible by adjusting a neighborhood size to the underlying structures of $\mu$ and $\Sigma$. Second, it may match the performance of the population Hotelling's $T^2$ test and thus is optimal in power under certain conditions. Third, since the neighborhood structure is explored by the linear regression, the test can be implemented without imposing restrictive structure assumptions on the unknown $\Sigma$. %and the other can possess better power under a local alternative condition. 
Last but not least, compared with tests via the random projection, the proposed test is easy to implement and computationally efficient. To select the optimal neighborhood size, we further propose a stability selection procedure to maximize the power via maximizing the signal-to-noise ratio. %The advantage of the proposed test is that by adaptively choose the neighborhood size, it can be more powerful than other existing tests. %Moreover, since the neighborhood structure is explored by the linear regression, the test can be implemented without imposing restrictive structure assumptions on unknown $\mu$ and $\Sigma$. %These two aspects enable the new test to surpass the capability of other existing tests and thus have a wide range of applications for real data analysis.   

%In this paper, we propose a new test statistic that is constructed by linking each component of the random vector $X_i$ in (\ref{model}) to its nearest predecessors through a linear regression model, and then formulated by targeting on either the sum-squared-based statistic or the $U$-statistic similar to Chen and Qin (2010). The asymptotic normality of the proposed test statistic is established under some mild conditions. It is shown that the power of the proposed test is boosted by exploiting data dependence. Most importantly, it maintains accurate type I error without assuming any specific structure of covariance matrix. These two aspects enable the new test to surpass the capability of other existing tests and thus have a wide range of applications in either genomic study or financial field. 

The rest of the paper is organized as follows. Section 2 defines {a class for $L_2$-norm statistics with a known $\Sigma$, in which the population Hotelling's $T^2$ test is shown to be optimal under certain conditions. %a class of oracle tests with a known $\Sigma$, and discusses their optimality in power under two different alternative conditions. 
The obtained results motivate us to formulate a Neighborhood-Assisted $T^2$ test statistic in Section 3, where we will investigate its asymptotic properties and also provide a stability procedure for choosing the optimal neighborhood size.} %Section 3 introduces the formulation of the Neighborhood-Assisted $T^2$ test statistic with the unknown $\Sigma$, investigates its asymptotic properties under some regularity conditions and also provides a stability procedure for choosing the optimal neighborhood size. 
Extensions of the Neighborhood-Assisted $T^2$ test to two-sample testing problem and non-Gaussian data are given in Section 4. Simulation and case studies are conducted in Sections 5 and 6 respectively, to demonstrate the empirical performance of the proposed test. A discussion about our results and other related work is given in Section 7. Technical proofs of main theorems are relegated to Appendix. Proofs of lemmas and propositions, and additional simulation results are included in a supplementary material.

\setcounter{section}{2} \setcounter{equation}{0}
\section*{\large 2. \bf Oracle Testing Procedures}

%\subsection*{2.1 \bf An Oracle Testing Procedure}

In the classical setting with $p$ fixed and $p \le n-1$, the Hotelling's $T^2$ test is uniformly most powerful within the class of statistics that are invariant with respect to the transformation $C X_i$ for a nonsingular matrix $C$. In high-dimensional setting, we consider a class of $L_2$-norm test statistics that are formulated similar to the Hotelling's $T^2$ statistic with a known $\Sigma$. The class consists of 
\be
T_0^2(\eta)=n \bar{X}^{\prime} \Sigma^{2\eta} \bar{X}, \label{Oracle}
\ee
where $\eta$ is a constant chosen from $(-\infty, \infty)$. 

The above class provides a unified framework to study the impact of data dependence via the covariance matrix $\Sigma$ on the power of the $L_2$-norm test. In particular, it includes three commonly used test statistics in the literature: the test statistic $T_0^2(0)$ without data transformation, the population Hotelling's $T^2$ test statistic $T_0^2(-1/2)$ with data transformed by $\Sigma^{-1/2}$, and the test statistic $T_0^2(-1)$ with data transformed by $\Sigma^{-1}$. Compared with the population Hotelling's $T^2$ statistic, the first one does not utilize any correlation information among variables and the corresponding test is expected to perform similar to the BS and CQ tests proposed in Bai and Saranadasa (1996), and Chen and Qin (2010), respectively. The last one has been identified as a superior transformation in higher criticism test (Hall and Jin, 2010), maximum test (Cai, Liu and Xia, 2014) and $L_2$ thresholding test (Chen, Li and Zhong, 2015). 

To explore the optimality within the class (\ref{Oracle}), we first establish the asymptotic normality of $T_0^2(\eta)$ under the following condition.      

\medskip
(C0). As $n \to \infty$, $p \to \infty$ and the matrix $\Sigma^{1+2\eta}$ satisfies that $\mbox{tr}\{(\Sigma^{1+2\eta})^4\}=o(\mbox{tr}^2\{(\Sigma^{1+2\eta})^2\})$.

The condition (C0) is similar to the one in Chen and Qin (2010). Rather than imposing any explicit relationship between $p$ and $n$, (C0) only requires a mild condition in terms of the covariance. It can be shown that (C0) is automatically satisfied if all the eigenvalues of $\Sigma$ are bounded.%away from infinity and zero.

\medskip

\noindent\textbf{Proposition 1.} {\it Under the model (\ref{model}) and (C0), as $n \to \infty$ and $p \to \infty$, }
$$\frac{T_{0}^2(\eta) - n\mu' \Sigma^{2\eta} \mu-\mbox{tr}(\Sigma^{1+2\eta})}{\sqrt{{2}\mbox{tr}(\Sigma^{2+4\eta})+{4}{n}\mu' \Sigma^{1+4 \eta} \mu}}  \xrightarrow{d} N(0, 1).$$

\medskip

Based on Proposition 1, a testing rule rejects the null hypothesis at the nominal significant level $\alpha$ if $T_0^2(\eta) \ge z_{\alpha} \sigma_0 (\eta)+\mbox{tr}(\Sigma^{1+2\eta})$, where $z_{\alpha}$ is the upper $\alpha$-quantile of $N(0, 1)$ and $\sigma_0(\eta)=\sqrt{2\mbox{tr}(\Sigma^{2+4\eta})}$. Furthermore, the power of the test is 
\[
\beta_{Or}(\mu, \eta)=\Phi\biggr\{-z_{\alpha}\frac{\sigma_0(\eta)}{\sigma(\eta)}+\frac{n\mu' \Sigma^{2\eta} \mu}{\sigma(\eta)} \biggr\},
\] 
where $\sigma(\eta)=\sqrt{{2}\mbox{tr}(\Sigma^{2+4\eta})+{4}{n}\mu' \Sigma^{1+4 \eta} \mu}$. Since $\sigma_0(\eta) \le \sigma(\eta)$, the power is determined by the signal-to-noise ratio 
\be
\mbox{SNR}_{Or}(\mu,\eta)=\frac{n\mu' \Sigma^{2\eta} \mu}{\sqrt{{2}\mbox{tr}(\Sigma^{2+4\eta})+{4}{n}\mu' \Sigma^{1+4 \eta} \mu}}, \label{SNR_oracle}
\ee
which depends on the value of $\eta$. The following proposition establishes the optimality of the population Hotelling's $T^2$ test under certain conditions. %study the optimal values of $\eta$ for the best power under two different scenarios. 

\medskip
\noindent\textbf{Proposition 2.} {\it Under the same conditions in Proposition 1, and the non-local alternative condition that $\mbox{tr}(\Sigma^{2+4\eta})=o(2n\mu' \Sigma^{1+4 \eta} \mu)$, the best power is attained by the population Hotelling's $T^2$ test with $\eta=-1/2$ in (\ref{Oracle}).}

\medskip

Note that if all the eigenvalues of $\Sigma$ are bounded, %away from $0$ and $\infty$, 
$\mbox{tr}(\Sigma^{2+4\eta})$ and $2n\mu' \Sigma^{1+4 \eta} \mu$ are at the orders of $p$ and $n\mu'\mu$, respectively.
The non-local alternative condition in Proposition 2 implies that $\mu'\mu \gg p / n$, which stands for a strong signal regime. Specially, with all components of $\mu$ being the same such that $\mu = \mu_{0}(1, \ldots, 1)'$, this condition requires $\mu_{0} \gg n^{-1/2}$. Proposition 2 states that the population Hotelling's $T^2$ test attains the best power under such a strong signal regime.  
%Proposition 2 states that under the fix alternative condition, the population Hotelling's $T^2$ test with $\Sigma^{-1/2}$ transformation attains the best power within the class (\ref{Oracle}). 

Different from the non-local alternative condition, the local alternative condition $2n\mu' \Sigma^{1+4 \eta} \mu=o\{\mbox{tr}(\Sigma^{2+4\eta})\}$ specifies a weak signal regime. Let $\lambda_{(1)} \leq \ldots \leq \lambda_{(p)}$ be the eigenvalues of $\Sigma$, and $\xi_{1}, \ldots, \xi_{p}$ be their corresponding orthonormal eigenvectors. Let $\bar{\lambda}^{2} = \sum_{j=1}^{p}\lambda_{(j)}^{2} / p$, $\tilde{\lambda}^{2} = \sum_{j=1}^{p}\lambda_{(j)}^{-2} / p$, and $S(\xi_{i_1}, \ldots, \xi_{i_k})$ be the linear space spanned by $\xi_{i_1}, \ldots, \xi_{i_k}$.
The following proposition compares the powers of three commonly discussed statistics $T_0^2(0)$, $T_0^2(-1/2)$ and $T_0^2(-1)$ under the local alternative condition. %transformations in $L_2$-norm test statistics.%As we will show next, under the local alternative condition, the power of the $L_2$-norm test within the class (\ref{Oracle}) can be dramatically affected by the underlying structure of $\mu$ and $\Sigma$. Therefore, there is no unique value of $\eta$ that leads to the best test in power. 

%To appreciate this, we specifically focus on three test statistics: the test statistic without data transformation ($\eta = 0$), the population Hotelling's $T^2$ test statistic with data transformed by $\Sigma^{-1/2}$ ($\eta = -1/2$), and the test statistic with data transformed by $\Sigma^{-1}$ ($\eta = -1$). Compared with the population Hotelling's $T^2$ statistic, the first one does not utilize any correlation information among variables, while the last one has been identified as a superior transformation in higher criticism test (Hall and Jin, 2010), maximum test (Cai, Liu and Xia, 2014) and $L_2$ thresholding test (Chen, Li and Zhong, 2015).

\medskip
{\bf Proposition 3.} {\it Let $m_{1} = \max\{j: \lambda_{(j)} \leq \min(\bar{\lambda}, \tilde{\lambda}^{-1})\}$, and $m_{2} = \min\{j: \lambda_{(j)} \geq \max(\bar{\lambda}, \tilde{\lambda}^{-1})\}$. Under the same conditions in Proposition 1, and the local alternative condition that $n\mu' \Sigma^{1+4 \eta} \mu=o\{\mbox{tr}(\Sigma^{2+4\eta})\}$,} the powers satisfy
\bea
\beta_{Or}(\mu, -1) \geq \beta_{Or}(\mu, -1/2) \geq \beta_{Or}(\mu, 0) &\mbox{\it if}& \mu \in S(\xi_{1}, \ldots, \xi_{m_1}), \nn \\
\beta_{Or}(\mu, -1) \leq \beta_{Or}(\mu, -1/2) \leq \beta_{Or}(\mu, 0) &\mbox{\it if}& \mu \in S(\xi_{m_2}, \ldots, \xi_{p}). \nn
\eea

%The power enhancement by transforming data via the precision matrix has been considered in Hall and Jin (2010) for the innovated higher criticism test, Cai, Liu and Xia (2014) in the max-norm based test and Chen, Li and Zhong (2014) in the test with thresholding and transformation. To the best of our knowledge, the results in Propositions 2 and 3 are the first ones that investigate the power performance of high-dimensional $L_2$-norm tests with test statistics defined in a general class (\ref{Oracle}). 
%\medskip

%Different from the transformation with $\eta = -1/2$ that leads to the optimal Hotelling's $T^2$ test under the non-local alternative, no universal optimal transformation exists for $L_2$-norm tests from the class (\ref{Oracle}) under the local alternative. 
Proposition 3 states that no universal optimal test exists in the class (\ref{Oracle}) under the local alternative. It also reveals the impact of the  structures of $\mu$ and $\Sigma$ on the power of $L_2$-norm tests in the weak signal regime. %Note that $S(\xi_{1}, \ldots, \xi_{m_1})$ and $S(\xi_{m_2}, \ldots, \xi_{p})$ are the spaces spanned by the eigenvectors associated with the $m_1$ smallest eigenvalues and the $p-m_2$ largest eigenvalues, respectively. Proposition 3 also reveals that the test based on $T_0^2(-1)$ possesses the best power among the three tests when the population mean $\mu$ lies in the space $S(\xi_{1}, \ldots, \xi_{m_1})$. However, the test without data transformation becomes the best one when $\mu$ lies in the space $S(\xi_{m_2}, \ldots, \xi_{p})$.
To put the results of Proposition 3 into an illustration, we consider the AR(1) model for $\Sigma$ satisfying $\sigma_{ij}=0.6^{|i-j|}$ for $1 \le i, j \le p$. The sample size $n=60$ and dimension $p=200$. We first design $\mu \in S(\xi_{m_2}, \ldots, \xi_{p})$ by a small cluster where only the first eight components of $\mu$ are non-zero with equal magnitude 0.2. A numerical computation gives that $\mbox{SNR}_{Or}(\mu, -1)=0.1448$, $\mbox{SNR}_{Or}(\mu, -1/2)=0.3826$ and $\mbox{SNR}_{Or}(\mu, 0)=0.5835$. On the contrary, we attain $\mu \in S(\xi_{1}, \ldots, \xi_{m_1})$ by randomly distributing the eight non-zero components from $\{1, \cdots, 200\}$. It gives a realization that $\mbox{SNR}_{Or}(\mu, -1)=1.8859$, $\mbox{SNR}_{Or}(\mu, -1/2)=1.7192$ and $\mbox{SNR}_{Or}(\mu, 0)=0.6304$.

%{\bf Remark 1.} With a known $\Sigma$, $T_0^2(0)$, $T_0^2(-1/2)$ and $T_0^2(-1)$ represent three $L_2$-norm test statistics based on original data, data transformed by $\Sigma^{-1/2}$, and data transformed by $\Sigma^{-1}$, respectively.  Propositions 2 and 3 specify the region for $\mu$ in which one test dominates the other two. Specially, under a strong signal regime defined by the non-local alternative condition, the population Hotelling's $T^2$ test is optimal for the best power. It thus deserves  thorough investigation in this paper.          

The established results in this section provide valuable insight into the impact of underlying structures of $\mu$ and $\Sigma$ on the power of the $L_2$-norm tests. Specially, under a strong signal regime specified by the non-local alternative condition, the population Hotelling's $T^2$ test is shown to be optimal for the best power. It therefore deserves to further extend its performance to the case of unknown $\Sigma$. In next section, we will propose a Neighborhood-Assisted Hotelling's $T^2$ test statistic and establish the corresponding asymptotic testing procedure. As we will see, the proposed test has the ability to adaptively choose a neighborhood size for the best power with respect to unknown structures of $\mu$ and $\Sigma$. Most importantly, it can match the population Hotelling's $T^2$ test under certain conditions and thus possesses the optimality.

%More precisely, if the neighborhood size is chosen to be zero, the proposed test statistic represents a standardized $T_0^2(0)$. When the neighborhood size is positive, its performance is close to the population Hotelling's $T^2$ test statistic $T_0^2(-1/2)$ %with $\eta=-1/2$ under certain conditions.       

%Subject to different conditions, Propositions 2 and 3 specify different $\eta$ for the oracle test to attain the best power. All the results in this section are established by assuming a known $\Sigma$. In next section, we will provide a Neighborhood-Assisted Hotelling's $T^2$ testing procedure, which can adaptively choose the neighborhood size for the best power with respect to unknown structures of $\mu$ and $\Sigma$. More precisely, if the neighborhood size is chosen to be zero, the proposed test represents a standardized oracle test with $\eta=0$. When the neighborhood size is positive, its performance is close to the population Hotelling's $T^2$ test $T_0^2(-1/2)$ %with $\eta=-1/2$ 
%under certain conditions. 

\setcounter{section}{3} \setcounter{equation}{0}
\section*{\large 3. \bf Data-Driven Testing Procedure}

\subsection*{3.1 \bf Test Statistic}

%The argument in last section suggests that we choose $\Sigma^{-1}$ in (\ref{Oracle}) to attain the best power for testing the hypotheses (\ref{eq:hypo1}) under some conditions. In practice, $\Sigma^{-1}$ is unknown and needs to be estimated. With estimated $\Sigma^{-1}$, one would expect to obtain a test statistic deserving a similar performance as $T_0^2(-1/2)$. However, the attempt to revive the Hotelling's $T^2$ by this way is hampered by two challenges. 

%First, it is very challenging to estimate $\Sigma^{-1}$ in high-dimensional setting unless $\Sigma^{-1}$ is presumably sparse. With some recent development, sparse $\Sigma^{-1}$ can be estimated by a regularized estimator $\hat{\Sigma}^{-1}$ through banding or thresholding (Bickel and Levina, 2008a, 2008b). However, in many real applications, $\Sigma^{-1}$ may not be sparse and thus its estimator cannot be obtained by the above methods. On the other hand, the optimality of the test established in Proposition 2 does not impose any structure assumption on $\Sigma^{-1}$. Therefore, reviving the Hotelling's $T^2$  without imposing too restrictive structure of $\Sigma^{-1}$ is very important from both practical and theoretical points of view. Second, even when the estimator $\hat{\Sigma}^{-1}$ of $\Sigma^{-1}$ with certain accuracy is available, small estimation errors of $\hat{\Sigma}^{-1}$ can be easily accumulated in $L_2$-norm statistic. This will lead to the failure of reviving $T^2$ by replacing $\Sigma^{-1}$ with $\hat{\Sigma}^{-1}$ (a detailed proof for this is given in Appendix).

The proposed test statistic for the hypotheses (\ref{eq:hypo1}) is  
\bea
T_{N}^2(k) &=& n\bar{X}^{\prime} \widehat{\Sigma}_{k}^{-1} \bar{X},  \label{newT2-1-NonCenter}
\eea
where, if $k=0$, $\widehat{\Sigma}_{0}^{-1}$ is a diagonal matrix with the $l$th diagonal element to be $(\sum_{i=1}^nX_{il}^2/n)^{-1}$ and if $k>0$, $\widehat{\Sigma}_{k}^{-1}$ is a banded Cholesky decomposition estimator obtained as follows. Let $\mathscr{X}\equiv (\mathscr{X}_1, \cdots, \mathscr{X}_p)=(X_1, \cdots, X_n)^{\prime}$ be the data matrix with $X_{i} = (X_{i1}, \cdots, X_{ip})^{\prime}$ defined in (\ref{model}) for $i=1, \cdots n$, and $\mathscr{X}_{l-k:l-1}$ be the columns from $l-k$ to $l-1$ of $\mathscr{X}$. Furthermore, let $\hat{A}$ be a lower triangular matrix with the nonzero elements in the $l$th row being $\hat{A}_{l, l-k:l-1}=(X_{1l}, \cdots, X_{nl})\mathscr{X}_{l-k:l-1}(\mathscr{X}_{l-k:l-1}^{\prime}\mathscr{X}_{l-k:l-1})^{-1}$, and $\widehat{D}$ be the diagonal matrix of $\hat{d}_{k, l}^2 = \mathscr{X}_{l}^{\prime} (I_n-H_l) \mathscr{X}_{l} / n$ with ${H}_{l} = \mathscr{X}_{l-k:l-1}(\mathscr{X}_{l-k:l-1}^{\prime}\\ \mathscr{X}_{l-k:l-1})^{-1}\mathscr{X}_{l-k:l-1}^{\prime}$ being the projection matrix spanned by $\mathscr{X}_{l-k:l-1}$. Thus, if $k>0$, $\widehat{\Sigma}_{k}^{-1}$ in (\ref{newT2-1-NonCenter}) is defined as
\be
\widehat{\Sigma}^{-1}_{k}=(I_p-\hat{A})^{\prime} \widehat{D}^{-1} (I_p-\hat{A}). \label{e_matNonCenter}
\ee

The advantages of proposing the test statistic $T_{N}^2(k)$ are as follows. First, it is closely related with a linear regression model, which enables us to analyze its asymptotic properties through the classical regression analysis. To see this, we choose a neighborhood size $k>0$  and regress ${\mathscr{X}}_l$ to its closest $k$ predecessors $({\mathscr{X}}_{l-k},\cdots, {\mathscr{X}}_{l-1})$ through the regression
\be
\mathscr{X}_{l} = \mathbf{1}\alpha_{l} + \mathscr{X}_{\max\{1, l-k\}:l-1}\gamma_{l} + \epsilon_{l}, \quad \mbox{for} \quad l = 2, \cdots, p,  \label{eq:RegInt}
\ee
where, based on (\ref{model}), the $k$-dimensional slope $\gamma_{l} = \Sigma_{\max\{1, l-k\}:l-1}^{-1}\Sigma_{\max\{1, l-k\}:l-1,l}$ and the intercept $\alpha_{l} = \mu_{l} - \mu^{\prime}_{\max\{1, l-k\}:l-1} \gamma_{l}$, the random error $\epsilon_{l} = (\epsilon_{1l}, \cdots, \epsilon_{nl})^{\prime}$ satisfying $\mbox{Var}(\epsilon_{il})=\mbox{Var}(X_{il}|\mathscr{X}_{\max\{1, l-k\}:l-1})\equiv d_{k,l}^2= \sigma_{ll}-\Sigma_{l, l-k:l-1} \Sigma_{l-k:l-1}^{-1} \Sigma_{l, l-k:l-1}^{\prime}$ for $i \in \{1, \cdots, n\}$, and the n-dimensional vector $\mathbf{1} = (1, \cdots, 1)^{\prime}$. Let $\hat{\epsilon}_{l} = (\hat{\epsilon}_{1 l}, \cdots, \hat{\epsilon}_{n l})^{\prime}$, $\hat{\alpha}_{l}$ and $\hat{\gamma}_{l}$ be the residuals, the least square estimators of $\alpha_{l}$ and $\gamma_{l}$, respectively. As shown in the supplementary material, the test statistic $T_{N}^2(k)$ can be written as  
\bea
T_{N}^2(k)\equiv \sum_{l=1}^p T_{N, l}^2 =\sum_{l=1}^p \frac{{ \mathcal{F}_{l}^{2}\hat{\alpha}_{l}^{2} }}{\hat{\epsilon}_{l}^{\prime}\hat{\epsilon}_{l}+{ \mathcal{F}_{l}\hat{\alpha}_{l}^{2} }}, \label{eq:CenterNonCenterRelation}
\eea
where $\mathcal{F}_{l} = \mathbf{1}^{\prime}(I - H_{l})\mathbf{1}$. Using the above expression, we will establish the asymptotic normality of $T_{N}^2(k)$ in Section 3.2 under some mild conditions.

Second, the neighborhood structure is adaptively explored by $T_{N}^2(k)$ through a neighborhood size $k$. If $k= 0$, $T_{N, l}^2$ in (\ref{eq:CenterNonCenterRelation}) is reduced to $n\bar{X}_l^2/\hat{\sigma}_{0l}^2$ for the sample mean of $l$th component $\bar{X}_l=n^{-1} \sum_{i=1}^n X_{il}$ and the sample variance under the null hypothesis $\hat{\sigma}_{0l}^2=n^{-1}\sum_{i=1}^n X_{il}^2$. 
The statistic $T_{N}^2(0)$ is a sum of $p$ standardized marginal statistics and its performance is close to the oracle test statistic $T_0^2(0)$ in (\ref{Oracle}) specially when all the diagonal covariances of $\Sigma$ are equal to 1. %$T_{N}^2(0)$ will be close to the oracle test statistic $T_0^2(0)$ in (\ref{Oracle}). 
On the other hand, if $k>0$, the non-diagonal matrix $\widehat{\Sigma}^{-1}_{k}$ is included in the test statistic. %Ideally, one expects to choose $k=p$ such that $\widehat{\Sigma}^{-1}_{p}$ is an estimate of $\Sigma^{-1}$ and $T_{N}^2(p)$ can perform similarly to the population Hotelling's $T^2$ statistic $T_0^2(-1/2)$. However, in the ``large $p$, small $n$" paradigm, $k$ should be chosen less than $n$ to avoid singularity of $\widehat{\Sigma}^{-1}_{k}$ in (\ref{e_matNonCenter}). Even though $k$ cannot be chosen to be $p$, 
We will demonstrate in next section that under some conditions, $T_{N}^2(k)$ %with a proper $k \ll p$ is still 
is able to perform as well as the population Hotelling's $T^2$ statistic $T_0^2(-1/2)$ and thus possesses certain optimality. To appreciate this, we first establish its asymptotic normality.    

%As mentioned above, the format of our NA-$T^2$ statistic is flexible as it provides a framework to unify the two oracle statistics in Section 2 by choosing $k=0$ and $k>0$, respectively. The optimal $k$ in $T_{NA}^2(k)$ should be chosen to maximize the power of the test. To obtain an explicit expression for the power, we first derive the asymptotic normality of $T_{NA}^2(k)$.   

\subsection*{3.2 \bf Asymptotic Results}

We first introduce some notation. Let $\Sigma_{l-k:l-1}$ and $\Sigma_{l, l-k:l-1}$ be the covariance matrix of $(X_{1\,l-k_{n}}, \cdots, X_{1\,l-1})^{\prime}$ and the covariance between $X_{1\,l}$ and $(X_{1\,l-k_{n}}, \cdots, X_{1\,l-1})^{\prime}$, respectively.
%To derive the asymptotic normality of $T_{N}^2(k)$, we first 
Define a population counterpart of $\widehat{\Sigma}^{-1}_{k}$ in (\ref{e_matNonCenter}) by     
\begin{eqnarray}
\Sigma_{k}^{-1}=(I_{p} - A_{k})^{\prime} D_{k}^{-1} (I_{p} - A_{k}), \label{Sigmak}
\end{eqnarray}
where $A_{k}$ is a $k$-banded lower triangular matrix with nonzero elements in the $l$th row equal to $\Sigma_{l, l-k:l-1} \Sigma_{l-k:l-1}^{-1}$, and $D_{k}$ is a diagonal matrix with the $l$th diagonal element
$d_{k,l}^2 = \sigma_{ll}-\Sigma_{l, l-k:l-1} \Sigma_{l-k:l-1}^{-1} \Sigma_{l, l-k:l-1}^{\prime}$. Specially, for $k=0$, $\Sigma_{0}^{-1}$ is a diagonal matrix with $l$th diagonal element equal to $\sigma_{ll}^{-1}$. %Moreover, we let 
%\be
%\sigma_{N}^{2} (k)= { 2 \operatorname{tr}\{(\Sigma_{k}^{-1} \Sigma)^2\} }+ 4n\mu^{\prime}\Sigma_{k}^{-1} \Sigma \Sigma_{k}^{-1} \mu. \label{stat-var}
%\ee

We need the following condition to establish the asymptotic normality of the proposed Neighborhood-Assisted $T^2$ test statistic (\ref{newT2-1-NonCenter}).

\medskip

(C1). As $n \to \infty$, $p \to \infty$ and $k=o(n)$ such that 
$$\mbox{tr}\{(\Sigma_{k}^{-1}\Sigma)^4\}=o(\mbox{tr}^2\{(\Sigma_{k}^{-1}\Sigma)^2\}).$$

%There exists a small positive constant $\varepsilon$ such that the eigenvalues of $\Sigma$ satisfy $\varepsilon < \lambda_{\min}(\Sigma) \leq \lambda_{\max}(\Sigma) < \varepsilon^{-1}$.

%(C4). As $n \to \infty$, $p \to \infty$ such that $p = o(n^2/k^4)$ and $k=o(n)$.% for $\phi < 2$ as $n 

\medskip

The condition (C1) describes the relationship among dimension $p$, neighborhood size $k$ and sample size $n$. The neighborhood size $k$ is chosen to be $o(n)$ to avoid singularity of $\widehat{\Sigma}^{-1}_{k}$ in (\ref{e_matNonCenter}). Similar to the condition (C0), (C1) does not impose any explicit relationship between $p$ and $n$, but instead requires a condition regarding the covariance matrices $\Sigma_{k}$ and $\Sigma$. A sufficient condition of $\mbox{tr}\{(\Sigma_{k}^{-1}\Sigma)^4\}=o(\mbox{tr}^2\{(\Sigma_{k}^{-1}\Sigma)^2\})$ is that all the eigenvalues of $\Sigma$ are bounded. %there exists a small positive constant $\varepsilon$ such that the eigenvalues of $\Sigma$ satisfy $\varepsilon < \lambda_{\min}(\Sigma) \leq \lambda_{\max}(\Sigma) < \varepsilon^{-1}$.\footnote{Can we provide the proof of this in SM? We may need more discussion toward condition (C0) and (C1) since both referee asked.} 
%However, it is more general to impose (C1) rather than require all the eigenvalues of $\Sigma$ bounded away from zero and infinity. For example, an equal correlation matrix $\Sigma$ considered in simulation studies of this paper, has one spiked eigenvalue that diverges to infinity as $p$ increases. By choosing $k \ge 2$, the spiked eigenvalue can be leveled off by $\Sigma_k^{-1}$ such that (C1) is satisfied (see Figure \ref{fig:size1} for an illustration).   
%As an illustration of (C1), we consider AR(1) and random sparse $\Sigma$ described in Section 5, respectively. 
As one illustration, we consider an AR(1) $\Sigma$. For $k \geq 0$, $\mbox{tr}\{(\Sigma_{k}^{-1}\Sigma)^4\} \sim p$ and $\mbox{tr}^2\{(\Sigma_{k}^{-1}\Sigma)^2\} \sim p^2$ and clearly (C1) holds, where for two real sequences $\{a_n\}$ and $\{b_n\}$, $a_n \sim b_n$ means they are at the same order. 
%Even with $k=0$, (C1) still holds as all the eigenvalues of $\Sigma$ are bounded.    

\medskip
 
\noindent\textbf{Theorem 1.} {\it Assume (\ref{model}) and (C1). Under $H_0$ of (\ref{eq:hypo1}), as $n \to \infty$,
%and each non-zero component of $\mu$ satisfying $\mu_i=o(1)$ for $k=0$ and $\mu_i=o(k^{-1/2})$ for $k>0$. 
%from the distribution $N(\mu, \Sigma)$.
\[
%\frac{T_{N}^2(k)-n\mu^{\prime}\Sigma_{k}^{-1}\mu-p}{\sigma_{N}(k)}\xrightarrow{d} N(0, 1),
\frac{T_{N}^2(k)-p}{\sigma_{N,0}(k)}\xrightarrow{d} N(0, 1),
\]
%where $\Sigma_{k}^{-1}$ and $\sigma_{N}(k)$ are defined in (\ref{Sigmak}) and (\ref{stat-var}), respectively.
where $\sigma_{N,0}^2(k)={ 2 }\operatorname{tr}\{(\Sigma_{k}^{-1} \Sigma)^2\}$ and $\Sigma_{k}^{-1}$ is defined by (\ref{Sigmak}).
Moreover, under $H_a$ of (\ref{eq:hypo1}), if $\sqrt{n}\mu_i=o\{(n/k)^{1/2}\}$ for $i=1, \cdots, p$, 
then as $n \to \infty$, 
\[
\frac{T_{N}^2(k)-n\mu^{\prime}\Sigma_{k}^{-1}\mu-p}{\sigma_{N}(k)}\xrightarrow{d} N(0, 1),
\]
where $\sigma_{N}^{2} (k)= { 2 \operatorname{tr}\{(\Sigma_{k}^{-1} \Sigma)^2\} }+ 4n\mu^{\prime}\Sigma_{k}^{-1} \Sigma \Sigma_{k}^{-1} \mu$.
} %is the special case of $\sigma_{N}(k)$ in (\ref{stat-var}) under $\mu = 0$.}
\medskip

%Specially, under the null hypothesis, $\{T_{N}^2(k)-p\}/\sigma_{N,0}(k)\xrightarrow{d} N(0, 1)$ where . 
%{\bf Theorem 1 requires no specific relationship between $n$ and $p$ as long as condition (C1) is satisfied.}

We require $\sqrt{n}\mu_i=o\{(n/k)^{1/2}\}$ under the alternative hypothesis in Theorem 1 in order to obtain the leading order expectation of $T_{N}^2(k)$ by Taylor expansion (see Lemma 3 in supplementary material for details). 
Note that $\sqrt{n}\mu_i$ could diverge to infinity due to $k=o(n)$.
The condition $\sqrt{n}\mu_i=o\{(n/k)^{1/2}\}$ spans a wide range of signal strength. Specially, if all components of $\mu$ are the same such that $\mu = \mu_{0}(1, \ldots, 1)'$, it implies the non-local alternative condition $\operatorname{tr}\{(\Sigma_{k}^{-1} \Sigma)^2\}=o(2n\mu^{\prime}\Sigma_{k}^{-1} \Sigma \Sigma_{k}^{-1} \mu)$ which specifies a strong signal regime. 
If $\sqrt{n}\mu_i \gg (n/k)^{1/2}$, the signal is so strong that it becomes an easier problem for any test.   

In order to implement a testing procedure, the unknown ${\sigma}_{N,0}^2(k)$ in Theorem 1 needs to be estimated. Similar to the estimator considered in Li and Chen (2012), $\sigma_{N,0}^2(k)$ can be estimated by
\begin{eqnarray}
\hat{\sigma}_{N,0}^2(k)&=& \frac{2\sum_{i \ne j}(X_i^{\prime} \widehat{\Sigma}^{-1}_{k}X_j)^2}{n(n-1)}
- \frac{4\sum_{i,j,k}^*X_i^{\prime} \widehat{\Sigma}^{-1}_{k} X_jX_j^{\prime} \widehat{\Sigma}^{-1}_{k}X_k}{n(n-1)(n-2)} \nonumber\\
&+& \frac{2\sum_{i,j,k,l}^*X_i^{\prime} \widehat{\Sigma}^{-1}_{k}X_j X_k^{\prime} \widehat{\Sigma}^{-1}_{k}X_l}{n(n-1)(n-2)(n-3)}, \label{var_est}
\end{eqnarray}
where $\sum^*$ denotes sum of mutually distinct indices and $\widehat{\Sigma}^{-1}_{k}$ is given by (\ref{e_matNonCenter}). The following theorem shows that $\hat{\sigma}_{N,0}^2(k)$ is a ratio-consistent estimator of ${\sigma}_{N,0}^2(k)$.

\medskip

\noindent\textbf{Theorem 2.} {\it Assume the same conditions in Theorem 1. As $n \to \infty$,} 
\[
\frac{\hat{\sigma}_{N,0}^2(k)}{{\sigma}_{N,0}^2(k)} \xrightarrow{p} 1. 
\]

Theorems 1 and 2 lead to a testing procedure that rejects $H_0$ in (\ref{eq:hypo1}) at the nominal significance level $\alpha$ if $T_{N}^2(k) \ge z_{\alpha}\hat{\sigma}_{N,0}(k)+p$  where $z_{\alpha}$ is the upper $\alpha$-quantile of $N(0, 1)$. Furthermore, 
%{\bf To investigate the power of the proposed test, we consider the case of weak signals under the alternative hypothesis such that  
%each non-zero component of $\mu$ satisfies $\mu_i=O(n^{-1/2})$ which converges to 0 as $n \to \infty$.
%Similar to Theorem 1, it can be shown that 
%$$\frac{T_{N}^2(k)-n\mu^{\prime}\Sigma_{k}^{-1}\mu-p}{\sigma_{N}(k)}\xrightarrow{d} N(0, 1)$$
%as $n\to\infty$ under the alternative hypothesis of (\ref{eq:hypo1}).
%Based on this asymptotic normality of $T_{N}^2(k)$, }
the power of the proposed test is 
\bea
\beta_{N}(\mu, k) &=& \Phi\biggl\{-z_{\alpha}\frac{\sigma_{N,0}(k)}{\sigma_{N}(k)} + \frac{n\mu^{\prime}{\Sigma}_{k}^{-1}\mu} {\sigma_{N}(k)}\biggr\}\{1 + o(1)\}. \nn
\eea
Since $\sigma_{N,0}(k) \le \sigma_{N}(k)$, the first term in $\Phi (\cdot)$ is bounded. The power is then determined by the signal-to-noise ratio
\be
\mbox{SNR}_{N}(\mu, k) :=\frac{n\mu^{\prime}{\Sigma}_{k}^{-1}\mu}
%{\sqrt{2 \mbox{tr}\{(\Sigma_{k}^{-1} \Sigma)^2\}+4n\mu^{\prime}\Sigma_{k}^{-1} \Sigma \Sigma_{k}^{-1} \mu}}. 
{ \sqrt{2\operatorname{tr}\{(\Sigma_{k}^{-1} \Sigma)^2\} +4n \mu^{\prime}\Sigma_{k}^{-1}\Sigma\Sigma_{k}^{-1} \mu} }. \label{SNR_new}
\ee

%Based on the derived signal-to-noise ratio, the following theorem that under a certain condition, the proposed NA-$T^2$ with a proper neighborhood size $k$ can mimic the performance of the population Hotelling's $T^2$ test, and thus enjoys certain optimality in power. specifies some conditions such that the proposed NA-$T^2$ with a proper neighborhood size $k$ can mimic the performance of $T_0^2(0)$ and $T_0^2(-1/2)$, respectively. 

The following theorem echoes the discussion at the end of Section 3.1 that under certain conditions, the proposed test has the flexibility to match performance of two competing tests by choosing different neighborhood size $k$.  

\medskip
{\bf Theorem 3.} {\it Assume the same conditions in Proposition 1 and Theorem 1. 
\begin{itemize}
\item[(1).] If the neighborhood size $k=0$, 
\be
\mbox{SNR}_{N}(\mu, 0)=\frac{n\mu' D^{-1} \mu}{\sqrt{{2}\mbox{tr}(R^{2})+{4}{n}\mu' D^{-1/2} R D^{-1/2} \mu}}, \label{SNR_diag}
\ee
where $D$ is the diagonal matrix formed by the diagonal elements $\{\sigma_{11}, \ldots, \sigma_{pp}\}$ of $\Sigma$, and $R = (r_{j_1j_2})$ is the correlation matrix such that $\Sigma = D^{1/2} R D^{1/2}$.

\item[(2).] By the Cholesky decomposition $\Sigma^{-1}=T(\Sigma)^{\prime}D^{-1}(\Sigma)T(\Sigma)$ with a lower triangular matrix $T(\Sigma)\equiv (t_{ij})$ and a diagonal matrix $D(\Sigma)$, we suppose that $\Sigma$ belongs to 
\begin{eqnarray}
\mathcal{V}^{-1}(\epsilon_0, \alpha, C)&=&\biggl\{\Sigma: 0<\epsilon_0\le \lambda_{\min}(\Sigma) \le \lambda_{\max}(\Sigma) \le \epsilon_0^{-1},\nonumber\\
&\qquad& \max_i \sum_{j<i-m}|t_{ij} |\le Cm^{-\alpha} \quad \mbox{for all} \quad m\le p-1 \biggr\}, \nonumber
\end{eqnarray}
%the power of the NA-$T^2$ test with the neighborhood size $k>0$ and $k=o(n)$ is asymptotically at the same order of that of the population Hotelling's $T^2$ test with $\eta=-1/2$.
then as $n, k \to \infty$, $\mbox{SNR}_{N}(\mu, k)=\mbox{SNR}_{Or}(\mu, -1/2)\{1+o(1)\}$, where $\mbox{SNR}_{Or}(\mu, -1/2)$ is the signal-to-noise ratio of the population Hotelling's $T^2$ test.
\end{itemize} }

\medskip

%Let $D_{0}$ be the diagonal matrix of $\{\sigma_{11}, \ldots, \sigma_{pp}\}$. Let $D_{0}^{1/2} = \mbox{diag}(\sqrt{\sigma_{11}}, \ldots, \sqrt{\sigma_{pp}})$. Notice that $\Sigma = D_{0}^{1/2} R D_{0}^{1/2}$, where $R = (r_{j_1j_2})$ is the correlation matrix.
%To better convey the merits of transformation, we consider the $L_2$ statistic on the transformed data $D^{-1/2}X_{i}$ such that\footnote{We need to introduce this statistic for our case of $k=0$.} 
%\be
%T^{2}_{d} = n\bar{X}' D^{-1} \bar{X}.
%\label{diag}\ee
%Following Proposition 1, it can be shown that the signal to noise ratio of the test statistic (\ref{diag}) is
%\be
%\mbox{SNR}_{d}=\frac{n\mu' D^{-1} \mu}{\sqrt{{2}\mbox{tr}(R^{2})+{4}{n}\mu' D^{-1/2} R D^{-1/2} \mu}}. \label{SNR_diag}
%\ee

%From Theorem 3 and (\ref{SNR_new}), the power of the proposed test can be shown to converge to 1 if $\mu^{\prime}\mu \geq C_0 p^{1/2} / n$ for a large positive constant $C_0$. 
The condition $\Sigma \in \mathcal{V}^{-1}(\epsilon_0, \alpha, C)$ specifies a  bandable structure of $\Sigma^{-1}$, which is commonly used in the literature (Bickel and Levina, 2008a). It holds for many structures including AR(1) and block diagonals given in Section 5. Theorem 3 states that with $k=0$, the proposed Neighborhood-Assisted $T^2$ statistic is equivalent to $T^{2}_{d} = n\bar{X}' D^{-1} \bar{X}$, which can be thought as the standardized oracle test statistic $T_{0}^{2}(0)$ in Section 2. On the other hand, under $\Sigma \in \mathcal{V}^{-1}(\epsilon_0, \alpha, C)$ and $k>0$, the proposed Neighborhood-Assisted $T^2$ test can match the population Hotelling's $T^2$ test. Such a structural flexibility demonstrates the advantage of the proposed Neighborhood-Assisted $T^2$ test: it can attain the optimality of the population Hotelling's $T^2$ test in the strong signal regime specified in Proposition 2 and obtain a better power between the population Hotelling's $T^2$ test and the test based on $T_{0}^{2}(0)$ in the weak signal regime discussed in Proposition 3.       

In real applications, both $\mu$ and $\Sigma$ are rarely known. It is therefore impossible for us to check the conditions in Propositions 2 and 3 to determine the best neighborhood size $k$ for the proposed test. Another challenge is that the bandable structure for real data may not be fulfilled. So the question of how to adaptively choose the neighborhood size $k$ to attain the best power possible with respect to unknown structures of $\mu$ and $\Sigma$ needs to be explored. In next section, we will provide a practical solution for such a question. %on the empirical choice of $k$ when there is no prior information about structures of $\mu$ and $\Sigma$. %From the analysis in the previous section, the optimal neighborhood size $k$ should be chosen to maximize the power of the test, which is equivalent to maximizing (\ref{SNR_new}) as the power is primarily determined by the signal-to-noise ratio.    

\subsection*{3.3 \bf Choosing the Neighborhood Size $k$}

%Theorem 3 and Proposition 4 in last section discuss the impact of the neighborhood size $k$ on the proposed Neighborhood-Assisted $T^2$ test from a theoretical point of view. In this section, we provide a practical solution on the empirical choice of $k$ when there is no prior information about structures of $\mu$ and $\Sigma$. From the analysis in the previous section, 
When there is no prior information about structures of $\mu$ and $\Sigma$, the proper neighborhood size $k$ should be chosen to maximize the power of the test, which is equivalent to maximizing (\ref{SNR_new}) as the power is primarily determined by the signal-to-noise ratio. In practice, the signal-to-noise ratio is unknown due to the unknown $\mu$ and $\Sigma$. At each $k$, we consider the following estimator for the unknown signal-to-noise ratio 
\be
\widehat{\mbox{SNR}}_{N}(k)=\frac{n\bar{X}^{\prime} \widehat{\Sigma}_{k}^{-1} \bar{X}-p}
{\sqrt{2n^{-2}\sum_{i \ne j}(X_i^{\prime} \widehat{\Sigma}^{-1}_{k}X_j)^2 +4n \hat{G}}}, \label{SNR_est}
\ee
where $\hat{G}=n^{-1}\sum_{i=1}^n \bar{X}^{\prime}\widehat{\Sigma}^{-1}_{k}(X_i-\bar{X})(X_i-\bar{X})^{\prime}\widehat{\Sigma}^{-1}_{k} \bar{X}-n^{-4}(\sum_{i\ne j}^n X_i^{\prime}\widehat{\Sigma}^{-1}_{k}X_j)^2$. The above estimator is constructed by plugging in the estimators of the numerator and denominator of the signal-to-noise ratio in (\ref{SNR_new}).    

Based on the proposed estimator for the signal-to-noise ratio, a stability selection procedure for the optimal neighborhood size is described as follows. 
\begin{itemize}
\item [] Step1: Given a sample of size $n$, choose a small set of integers $\{0, 1, \cdots, m\}$ with $m \ll n$, say $m=n/10$, for the possible values of $k$ to meet the assumption that $k=o(n)$. 
\item[] Step 2: Randomly divide the sample into $H$ parts with equal size. 
\item [] Step 3: Drop the $h$th ($h=1, \cdots, H$) part, and use the remaining $H-1$ parts of the sample to select the neighborhood size $\hat{k}_h$ through the criterion 
\[
\hat{k}_h=\argmax_{k \in \{0,1, \cdots, m\}} \widehat{\mbox{SNR}}_{N}(k),
\]
where $\widehat{\mbox{SNR}}_{N}(k)$ is defined by (\ref{SNR_est}). 
\item [] Step 4: Repeat Step 3 for all $h=1, \cdots, H$ to obtain $\{\hat{k}_1, \cdots, \hat{k}_H \}$. The optimal neighborhood size is defined to be the median of $\{\hat{k}_1, \cdots, \hat{k}_H \}$. 
 \end{itemize}

Our empirical study suggests to choose $H$ between 4 and 10 in Step 2 for satisfactory performance of the stability selection procedure. In Section 5, we will evaluate the numerical performance of the above procedure for optimal neighborhood size. As we will see, the stability selection procedure is able to choose the optimal neighborhood size with respect to different structures of $\mu$ and $\Sigma$, such that the testing procedure can attain its best power.

\setcounter{section}{4} \setcounter{equation}{0}
\section*{\large 4. \bf Some Extensions}

\subsection*{4.1 \bf Two-Sample Case}

The proposed Neighborhood-Assisted $T^2$ test for the one-sample mean scenario can be extended to the inference of the two-sample comparison. Suppose 
\begin{eqnarray}
X_{ij}=\mu_i+\epsilon_{ij}, \,\, \epsilon_{ij}\stackrel{i.i.d.}{\sim}\mbox{N}(0, \Sigma_i)\, \, \, \mbox{for} \,\,i=1, 2 \,\,\, \mbox{and} \, \,\, 1\le j\le n_i, \label{model1}
\end{eqnarray}
where $\mu_i$ is a $p$-dimensional population mean vector and $\Sigma_i$ is a $p\times p$ covariance matrix. Let $\delta=\mu_1-\mu_2=(\delta_1,\cdots,\delta_p)^{\prime}$. We consider to test
\be
H_{0}^* : \delta= 0  \quad  \mbox{vs} \quad H_{1}^* : \delta
\ne 0. \label{eq:hypo2}
\ee
%As demonstrated in this paper,

To extend the proposed Neighborhood-Assisted $T^2$ test for the two-sample hypotheses (\ref{eq:hypo2}), we convert this two-sample inference problem into a one-sample problem. A similar idea has been proposed in Anderson (2003) to construct the two-sample HotellingÕs $T^2$ test statistic with unequal covariance matrices. Without loss of generality, we assume $n_1 \le n_2$. Let
\be
Y_i= X_{1i}-\sqrt{\frac{n_1}{n_2}} X_{2i}+\frac{1}{\sqrt{n_1n_2}}\sum_{j=1}^{n_1}X_{2j}-\frac{1}{n_2}\sum_{l=1}^{n_2}X_{2l}, \qquad i=1,\cdots, n_1. \label{var_t}
\ee
Under the model (\ref{model1}), $Y_i \stackrel{i.i.d.}{\sim} \mbox{N}(\delta, \Sigma_w)$ for $i=1,\cdots, n_1$ with $\Sigma_w=\Sigma_1+\frac{n_1}{n_2}\Sigma_2$. By the transformation (\ref{var_t}), the average of the transformed variables $\bar{Y}=\frac{1}{n_1} Y_i=\bar{X}_1-\bar{X}_2$ is a natural estimator for $\delta$. 

A two-sample Neighborhood-Assisted $T^2$ test statistic for the hypotheses (\ref{eq:hypo2}) can be constructed by replacing $X_i$ with $Y_i$ in (\ref{newT2-1-NonCenter}) as
\be
T_{N}^2(k)=n_1 \bar{Y}^{\prime} \hat{\Sigma}_{w,k}^{-1} \bar{Y}, \label{newT2-1-2}
\ee
where the two-sample estimator $\hat{\Sigma}_{w,k}^{-1}$ can be obtained via replacing $X_i$ by $Y_i$ in (\ref{e_matNonCenter}). The established asymptotic results for the one-sample Neighborhood-Assisted $T^2$ can be directly applied to the above two-sample Neighborhood-Assisted $T^2$ test statistic.  

%One concern of (\ref{var_t}) is that some observations in the larger sample might be sacrificed for constructing $\hat{\Sigma}_{w,k}^{-1}$. However, as shown in Anderson (2003), this sacrifice is not very important.  %Also a two-sample U statistic analogous to (\ref{newT2-u}) is
%\be
%K_U^2=\frac{1}{n_1(n_1-1)}\sum_{i \ne j}^{n_1} Y_i^{\prime} \hat{\Sigma}_{w,k}^{-1}Y_j. \label{newT2-u-2}
%\ee

%With (\ref{var_t}), the combined two-sample NA-$T^2$ test statistic is obtained by replacing $X$ by $Y$ in (\ref{eq:combine}). Most importantly, the established asymptotic properties in Theorem 4 can be directly applied to the combined two-sample NA-$T^2$ test.  

\subsection*{4.2 \bf Non-Gaussian Data}

The results for the proposed Neighborhood-Assisted $T^2$ test can also be extended to non-Gaussian data. Instead of the Gaussian assumption in (\ref{model}), we consider $X_{i} = (X_{i1}, \cdots, X_{ip})^{\prime}$ to be i.i.d. from a distribution with mean $\mu$ and covariance $\Sigma = (\sigma_{j_1j_2})_{p \times p}$. To establish the asymptotic normality of the Neighborhood-Assisted $T^2$ test statistic under non-Gaussian data, we need two additional conditions.     

(C2). For any $j = 1, \cdots, p$, ${\E}\big\{\exp(q X_{1j})\big\} < \infty$ if $|q| < Q$ and $Q$ is a positive constant.

(C3). (i) $X_i=\Gamma Z_i + \mu$, where $\Gamma$ is a $p\times m$  matrix of constants with
$m\ge p$, $\Gamma \Gamma' = \Sigma$,  and $Z_1$, $\cdots$, $Z_n$ are
IID $m$-dimensional random vectors such that ${\E}(Z_1)=0$ and
${\V}(Z_1)=I_{m}$. (ii) For $Z_1=(Z_{1 1}, \dots, Z_{1 m})^T$,
$\{Z_{1 l}\}_{l=1}^{m}$ are independent with uniformly bounded
$8$-th moment, and there exist finite constants $\Delta$ and
$\omega$ such that ${\E}(z_{1 l}^4)=3+\Delta$ and  ${\E}(z_{1
l}^3)=\omega$ for $l=1, \cdots, m$.

Condition (C2) assumes existence of moment generating functions of data, which is required for the results of moderate and large deviation. Condition (C3) prescribes higher moments of $X_{i}$ without assuming any parametric distribution of data. Note that Gaussian distribution satisfies both those conditions. Our preliminary investigation has shown that the asymptotic normality of the Neighborhood-Assisted $T^2$ test statistic can be established under (C2) and (C3), and a more restrictive condition that $p=o(n^2)$. The question of whether the restrictive condition between $p$ and $n$ can be relaxed deserves further investigation. We will leave this to future study. 

%\noindent\textbf{Theorem 6.} {\it Under $H_{0}$ of (\ref{eq:hypo1}) and Conditions (C1) - (C4), as $n \to \infty$, 
%$${U_*^2} \xrightarrow{d} N(0, 1), $$ }
%where $U_*^2$ is given by (\ref{eq:combine}).
%\medskip

\setcounter{section}{5} \setcounter{equation}{0}
\section*{\large 5. \bf Simulation Studies}

Simulation studies were conducted to evaluate the empirical performance of the proposed Neighborhood-Assisted $T^2$ test. For comparison purpose, we also considered the one-sample CQ, BS and SD tests, proposed by Chen and Qin (2010), Bai and Sarandasa (1996) and Srivastava and Du (2008), respectively.  These tests were included because they have similar power performance as the oracle test with $\eta=0$ in Section 2. %More precisely, the one-sample CQ test statistic
%\[
%T_{CQ}^2=\frac{1}{n(n-1)}\sum_{i \ne j}^n X_i^{\prime} X_j.
%\]
%The BS test statistic proposed by Bai and Sarandasa (1996) is
%\[
%T_{BS}^2=\frac{n\bar{X}^{\prime}\bar{X}-\mbox{tr}(S_n)}{\sqrt{\frac{2n(n-1)}{(n+1)(n-2)}\{\mbox{tr}(S_n^2)-\frac{1}{n-1}\mbox{tr}^2(S_n) \}}}.
%\]
%The SD test proposed by Srivastava and Du (2008) has the test statistic
%\[
%T_{SD}^2=\frac{n\bar{X}^{\prime}\{\mbox{diag}(S_n)\}^{-1}\bar{X}-\frac{(n-1)p}{n-3}}{\sqrt{2\{\mbox{tr}(R^2)-\frac{p^2}{n-1}\}\{1+\frac{\mbox{tr}(R^2)}{p^{3/2}}\}}},
%\]
%where $\mbox{diag}(S_n)$ is the diagonal matrix formed by the diagonal elements of $S_n$ and $R=\mbox{diag}(S_n)^{-1/2}S\mbox{diag}(S_n)^{-1/2}$. The asymptotic normality of these test statistics have been established in the literature. 
Another test we compared is the oracle test with the test statistic $T_0^2(-1/2)$. Its asymptotic normality is established in Proposition 1. %Our simulation studies will confirm the theoretical findings that the proposed Neighborhood-Assisted $T^2$ test is able to adjust its neighborhood size to attain its best power.  %these tests when they possess better power. Again, our simulation studies will confirm that the proposed Neighborhood-Assisted $T^2$ test can match this oracle test when it attains the best power.

The data were simulated from $N(\mu, \Sigma)$. Under $H_0$, we assigned $\mu=0$. Under $H_1$, $\mu$ had $[p^{1-\beta}]$ non-zero entires, where $[a ]$ denotes the integer part of $a$. We considered both random signals in the sense that $[p^{1-\beta}]$ non-zero entires were randomly selected from $\{1,\cdots, p\}$, and clustering signals allocated in the first $[p^{1-\beta}]$ components. The value of each non-zero entry was $r$. The two parameters $\beta>0$ and $r>0$ were chosen to control the sparsity and strength of signals, respectively. The sample size was chosen to be $n=60$. The dimensions of random vector were $p= 200, 400$ and $1000$, respectively. All the simulation results were based on $1000$ replications with nominal significant level $\alpha=0.05$.

To model data dependence, we considered the following patterns of the covariance matrix $\Sigma=(\sigma_{ij})$:
\begin{itemize}
\item[ ] (a). AR(1) model:  $\sigma_{ij}=0.6^{|i-j|}$ for $1\le i, j\le p$.
\item[ ] (b). Block diagonal model: $\sigma_{ii}=1$ for $i=1,\cdots,p$, and $\sigma_{ij}=0.6$ for $2(k-1)+1\le i\ne j \le 2k$ where $k=1, \cdots, 4$.
\item[ ] (c). Random sparse matrix model: first generate a $p \times p$ matrix $\Gamma$ each row of which has only four non-zero element that is randomly chosen from $\{1, \cdots, p\}$ with magnitude generated from Unif(1, 2) multiplied by a random sign. Then $\Sigma=\Gamma \Gamma^{T}+\bm{\mbox{I}}$ where $\bm{\mbox{I}}$ is a $p \times p$ identity matrix.
\item[ ] (d). Equal correlation setting: $\sigma_{ii}=1$ for $i=1,\cdots,p$, $\sigma_{ij}=0.6$ for $i \ne j$.
\end{itemize}
Clearly, models (a) and (b) specify the bandable structure of $\Sigma$. Model (c) leads to a sparse structure of $\Sigma$ that is different from those in models (a) and (b). And model (d) stands for a structure beyond commonly assumed sparse or bandable pattern, which violates some key assumptions in the CQ, BS and SD tests.  For example, the CQ test requires the covariance matrix to satisfy $\mbox{tr}(\Sigma^4)=o\{\mbox{tr}^2(\Sigma^2)\}$ that is not held because the equal correlation matrix in model (d) has one unbounded eigenvalue that diverges as dimension $p$ increases. As a result, the tests of CQ, BS and SD are hampered by such covariance structure.

Different from the CQ, BS and SD tests, the proposed Neighborhood-Assisted $T^2$ test is still able to be implemented with the covariance structure specified in model (d). To appreciate this, we notice that our proposed Neighborhood-Assisted $T^2$ test statistic incorporates data dependence by the regularized estimator through banding Cholesky factor, which is an estimator of $\Sigma_{k}^{-1}$ defined by (\ref{Sigmak}). As demonstrated in Figure \ref{fig:size1}, there exists a $0<k=o(n)$ such that the spiked eigenvalue of $\Sigma$ in model (d) is leveled off by $\Sigma_{k}^{-1}$ to make the condition (C1) satisfied. %Since $\Sigma_{k}^{-1}\Sigma \to I_p$ as $k \to p$\footnote{We need to avoid $k \to p$}, it can be shown that there exists a $0<k=o(n)$ such that the spiked eigenvalue of $\Sigma$ in model (d) can be leveled off by $\Sigma_{k}^{-1}$ and the condition (C1) is satisfied (see Figure \ref{fig:size1} for more details).   
%As a result, the proposed residual tests can still be implemented for testing the hypotheses (\ref{eq:hypo1}) even though the CQ, BS and SD tests cannot be.

\subsection*{5.1 \bf The Stability Selection Procedure for Optimal Neighborhood Size $k$}

The Neighborhood-Assisted $T^2$ test depends on the choice of the neighborhood $k$. In Section 3.3, we introduced a stability selection procedure for the optimal $k$ such that the Neighborhood-Assisted $T^2$ test can attain its best power. Here, we demonstrate its performance via some numerical studies. The data were generated by the same setup at the beginning of Section 5, with sample size $n=60$, dimension $p=200$, $\beta=0.6$ and $r=0.2$. To meet the requirement that $k=o(n)$ in (C1), we restricted $k \in \{0, 1, \cdots, 10\}$. Again, all the results were based on $1000$ replicates. %In particular, we demonstrate the robustness of the procedure with respect to randomly distributed entries and clustering entries of $\mu$, respectively, and different structures of $\Sigma$. 

\begin{figure}[tbh!]
\centering
  \includegraphics[width=13cm]{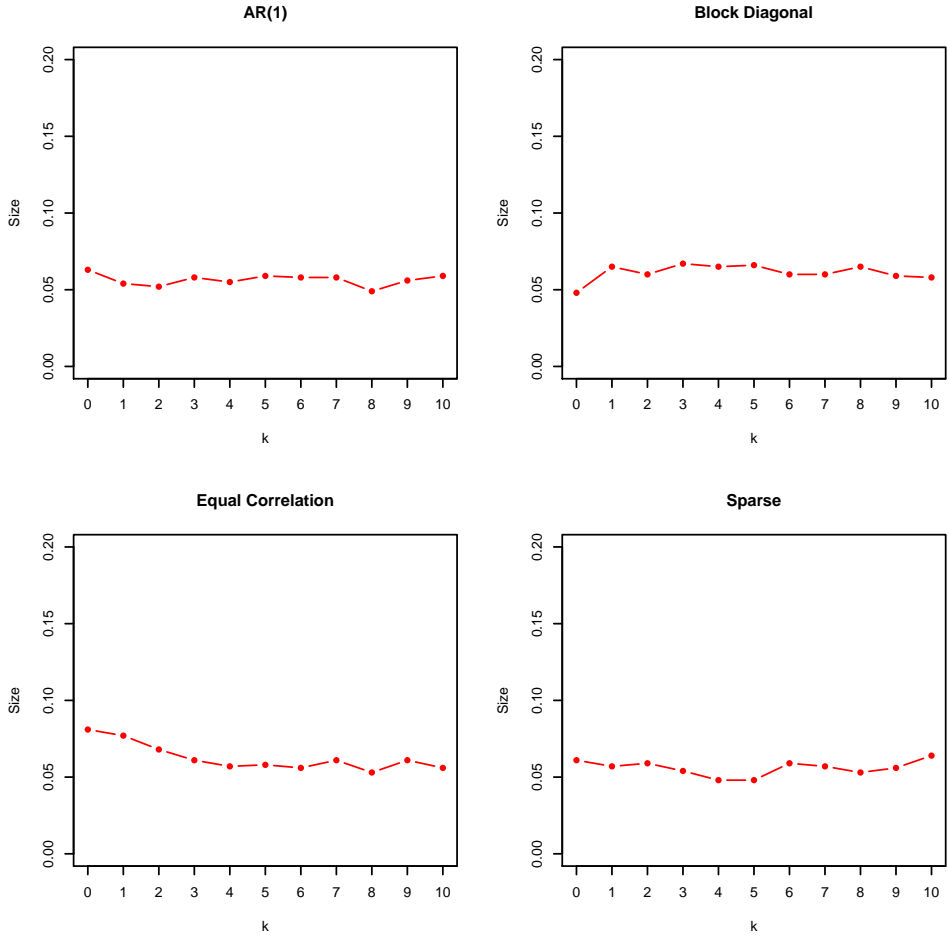}
  \caption{Empirical sizes of the Neighborhood-Assisted $T^2$ tests with neighborhood sizes chosen from $\{0, 1, \cdots, 10\}$, subject to different structures of $\Sigma$.}
  \label{fig:size1}
\end{figure}

According to Theorem 1, the size of the Neighborhood-Assisted $T^2$ test is not  affected by the choice of $k$ under (C1). Figure \ref{fig:size1} confirms this by the empirical sizes of the Neighborhood-Assisted $T^2$ test with different neighborhood sizes subject to different structures of $\Sigma$. Clearly, all sizes were close to the nominal significant level $0.05$ except choosing $k=0, 1, 2$ for the equal correlation structure of $\Sigma$. It happened because the spiked eigenvalue in $\Sigma$ cannot be leveled off by choosing $k=0, 1, 2$. However, as $k \ge 3$, its effect can be attenuated such that (C1) is satisfied.

\begin{figure}[tbh!]
\centering
  \includegraphics[width=13cm]{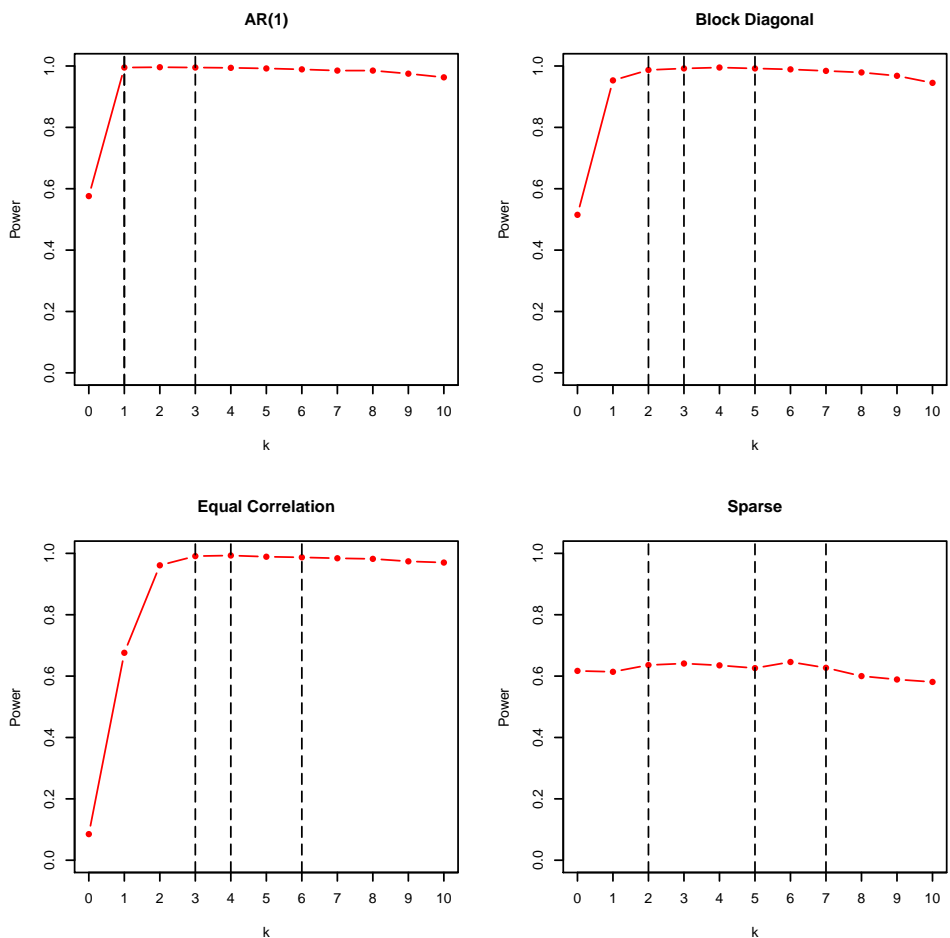}
  \caption{Empirical powers of the Neighborhood-Assisted $T^2$ tests with neighborhood sizes chosen from $\{0, 1, \cdots, 10\}$, subject to randomly distributed $\mu$ and different structures of $\Sigma$. Three vertical dash lines represent the first quantile, median and third quantile of the $1000$ selected optimal neighborhood sizes by the proposed stability neighborhood selection procedure.}
  \label{fig:Random2}
\end{figure}

\begin{figure}[tbh!]
\centering
  \includegraphics[width=13cm]{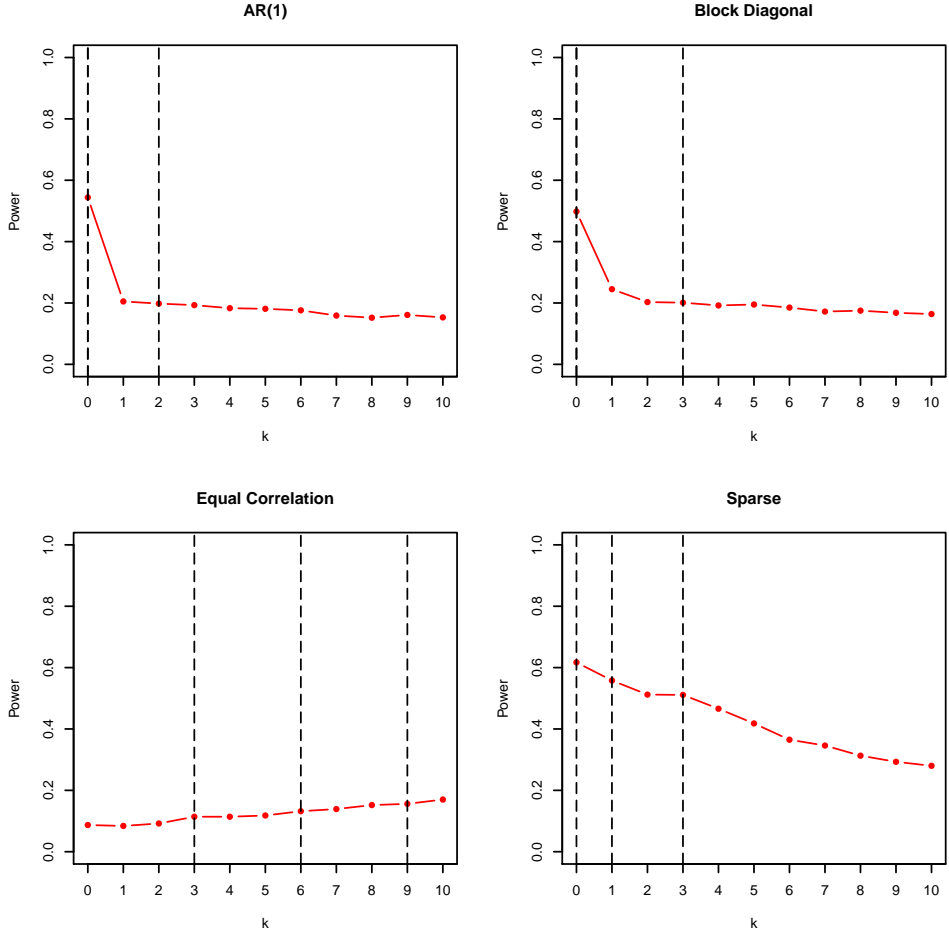}
  \caption{Empirical powers of the Neighborhood-Assisted $T^2$ tests with neighborhood sizes chosen from $\{0, 1, \cdots, 10\}$, subject to clustering $\mu$ and different structures of $\Sigma$. Three vertical dash lines represent the first quantile, median and third quantile of the $1000$ selected optimal neighborhood sizes by the proposed stability neighborhood selection procedure.}
  \label{fig:Cluster3}
\end{figure}

Unlike the size, the power of the Neighborhood-Assisted $T^2$ test depends on the choice of $k$. Figures \ref{fig:Random2} and \ref{fig:Cluster3} confirm this by the empirical powers of the Neighborhood-Assisted $T^2$ test under different neighborhood sizes, subject to randomly distributed and clustering $\mu$, respectively, and different structures of $\Sigma$. %It can be seen that the empirical powers of the test were affected by the choice of $k$. 
Moreover, the optimal $k$ that maximizes the empirical powers varied with the structures of $\mu$ and $\Sigma$. To evaluate the proposed stability neighborhood selection procedure in Section 3.3, we obtained the optimal neighborhood size for each of $1000$ replicates with the number of randomly split subsamples $H=5$. Three vertical dash lines in Figures \ref{fig:Random2} and \ref{fig:Cluster3}, represent the first quantile, median and third quantile of the $1000$ selected optimal neighborhood sizes. Clearly, all three vertical dash lines were close to the location where the maximal power was attained, demonstrating the satisfactory performance of the proposed optimal neighborhood selection procedure.

  \subsection*{5.2 \bf Size and Power Performance}

Table \ref{table3} displays the empirical sizes of all five tests considered in this paper. Since the size is not affected by the neighborhood size as long as $k=o(n)$, we chose $k=3$ for the Neighborhood-Assisted $T^2$ test under the null hypothesis. With the bandable and sparse covariance matrices specified in models (a), (b) and (c), all the tests had the empirical sizes close to the nominal level of significance $\alpha=0.05$. With the equal correlation matrix defined in model (d), a key condition in the CQ, BS and SD tests cannot be fulfilled, which leads to size distortion of those tests as shown in Table \ref{table3}. Different from the CQ, BS and SD tests, the proposed Neighborhood-Assisted $T^2$ test was still able to maintain very accurate sizes with $k=3$. %Again, the size distortion of the c-NA-$T^2$ test was mitigated by the comb-NA-$T^2$ test. % that has size performance similar to the NC-Res.

\begin{table}[t]
\tabcolsep 3pt
\centering
\caption{Empirical sizes of Chen and Qin's test (CQ), Bai and Sarandasa's test (BS), Srivastava and Du's test (SD), the Oracle test, and the proposed Neighborhood-Assisted $T^2$ test (New) subject to different structures of $\Sigma$.}
\label{table3}
\begin{tabular}{|l|ccc|ccc|ccc|ccc|}
\hline  &\multicolumn{3}{c|}{Model (a)}&\multicolumn{3}{c|}{Model (b)} &\multicolumn{3}{c|}{Model (c)} &\multicolumn{3}{c|}{Model (d)} \\[1mm]
\cline{1-4} \cline{5-7} \cline{8-10} \cline{11-13}& $200$& $400$& $1000$& $200$& $400$ &$1000$& $200$& $400$ &$1000$ & $200$& $400$ &$1000$ \\
\hline
CQ & $0.055$ & $0.064$& $0.069$ &$0.046$ & $0.052$& $0.055$ &$0.054$ & $0.045$& $0.063$ &$0.086$ & $0.062$& $0.069$\\[1 mm]
BS & $0.055$ & $0.064$& $0.069$ &$0.047$ & $0.052$& $0.055$&$0.054$ & $0.044$& $0.063$ &$0.085$ & $0.062$& $0.069$\\[1 mm]
SD & $0.051$ & $0.041$& $0.034$ &$0.037$ & $0.031$& $0.025$ &$0.043$ & $0.028$& $0.029$ &$0.017$ & $0.007$& $0.004$\\[1 mm]
Oracle & $0.052$ & $0.048$& $0.054$&$0.049$ & $0.049$& $0.049$ &$0.052$ & $0.053$& $0.046$&$0.044$ & $0.045$& $0.067$\\[1 mm]
%c-NA & $0.038$ & $0.037$& $0.042$ &$0.036$ & $0.026$& $0.032$ &$0.086$ & $0.083$& $0.084$ &$0.082$ & $0.064$& $0.090$\\[1 mm]
%NC-Res & $0.047$ & $0.051$& $0.067$ &$0.054$ & $0.041$& $0.056$ &$0.051$ & $0.057$& $0.061$ &$0.053$ & $0.045$& $0.066$
%\\[1 mm]
New & $0.047$ & $0.051$& $0.067$ &$0.054$ & $0.041$& $0.056$ &$0.051$ & $0.057$& $0.061$ &$0.053$ & $0.045$& $0.066$
\\[1 mm]
\hline
\end{tabular}
\end{table}

\begin{table}[tbh!]
\tabcolsep 4pt
\centering
\caption{Empirical powers of Chen and Qin's test (CQ), Bai and Sarandasa's test (BS), Srivastava and Du's test (SD), the Oracle test, and the proposed Neighborhood-Assisted $T^2$ test (New) with randomly distributed $\mu$ and $\Sigma$ specified in model (a).}
\label{table4}
\begin{tabular}{lccccccccccc}
\hline \multicolumn{12}{c}{$\beta=0.4$}                                                                  \\ \hline
         & \multicolumn{3}{c}{$p=200$} &  & \multicolumn{3}{c}{400} &  & \multicolumn{3}{c}{1000} \\ \cline{2-4} \cline{6-8} \cline{10-12}
         & $r=0.2$   & 0.4    & 0.6    &  & $r=0.2$ & 0.4   & 0.6   &  & $r=0.2$  & 0.4   & 0.6   \\ \hline
CQ       & 0.575     & 1      & 1      &  & 0.624   & 1     & 1     &  & 0.723    & 1     & 1     \\
BS       & 0.575     & 1      & 1      &  & 0.625   & 1     & 1     &  & 0.723    & 1     & 1     \\
SD       & 0.525     & 1      & 1      &  & 0.540   & 1     & 1     &  & 0.604    & 1     & 1     \\
Oracle   & 0.996     & 1      & 1      &  & 1       & 1     & 1     &  & 1        & 1     & 1     \\
New & 0.995     & 1      & 1      &  & 1       & 1     & 1     &  & 1        & 1     & 1     \\ \hline
\multicolumn{12}{c}{$\beta=0.8$}                                                                  \\ \hline
         & \multicolumn{3}{c}{$p=200$} &  & \multicolumn{3}{c}{400} &  & \multicolumn{3}{c}{1000} \\ \cline{2-4} \cline{6-8} \cline{10-12}
         & $r=0.2$   & 0.4    & 0.6    &  & $r=0.2$ & 0.4   & 0.6   &  & $r=0.2$  & 0.4   & 0.6   \\ \hline
CQ       & 0.104     & 0.255  & 0.673  &  & 0.058   & 0.184 & 0.458 &  & 0.094    & 0.136 & 0.360 \\
BS       & 0.105     & 0.254  & 0.673  &  & 0.058   & 0.183 & 0.459 &  & 0.094    & 0.136 & 0.360 \\
SD       & 0.075     & 0.212  & 0.618  &  & 0.042   & 0.139 & 0.387 &  & 0.063    & 0.087 & 0.265 \\
Oracle   & 0.205     & 0.872  & 1      &  & 0.136   & 0.656 & 0.995 &  & 0.126    & 0.559 & 0.983 \\
New & 0.162     & 0.724  & 0.992  &  & 0.113   & 0.503 & 0.878 &  & 0.128    & 0.381 & 0.746 \\ \hline
\end{tabular}
\end{table}

\begin{table}[tbh!]
\tabcolsep 4pt
\centering
\caption{Empirical powers of Chen and Qin's test (CQ), Bai and Sarandasa's test (BS), Srivastava and Du's test (SD), the Oracle test, and the proposed Neighborhood-Assisted $T^2$ test (New) with randomly distributed $\mu$ and $\Sigma$ specified in model (b).}
\label{table5}
\begin{tabular}{cccclccclccc}
\hline
\multicolumn{12}{c}{$\beta=0.4$}                                                                  \\ \hline
         & \multicolumn{3}{c}{$p=200$} &  & \multicolumn{3}{c}{400} &  & \multicolumn{3}{c}{1000} \\ \cline{2-4} \cline{6-8} \cline{10-12}
         & $r=0.2$   & 0.4    & 0.6    &  & $r=0.2$ & 0.4   & 0.6   &  & $r=0.2$  & 0.4   & 0.6   \\ \hline
CQ       & 0.534     & 1      & 1      &  & 0.593   & 1     & 1     &  & 0.659    & 1     & 1     \\
BS       & 0.534     & 1      & 1      &  & 0.592   & 1     & 1     &  & 0.658    & 1     & 1     \\
SD       & 0.463     & 1      & 1      &  & 0.518   & 1     & 1     &  & 0.536    & 1     & 1     \\
Oracle   & 0.993     & 1      & 1      &  & 0.999   & 1     & 1     &  & 1        & 1     & 1     \\
New & 0.979     & 1      & 1      &  & 0.997   & 1     & 1     &  & 0.999    & 1     & 1     \\ \hline
\multicolumn{12}{c}{$\beta=0.8$}                                                                  \\ \hline
         & \multicolumn{3}{c}{$p=200$} &  & \multicolumn{3}{c}{400} &  & \multicolumn{3}{c}{1000} \\ \cline{2-4} \cline{6-8} \cline{10-12}
         & $r=0.2$   & 0.4    & 0.6    &  & $r=0.2$ & 0.4   & 0.6   &  & $r=0.2$  & 0.4   & 0.6   \\ \hline
CQ       & 0.083     & 0.228  & 0.615  &  & 0.090   & 0.178 & 0.426 &  & 0.070    & 0.137 & 0.342 \\
BS       & 0.082     & 0.228  & 0.614  &  & 0.091   & 0.178 & 0.426 &  & 0.071    & 0.137 & 0.342 \\
SD       & 0.063     & 0.182  & 0.542  &  & 0.059   & 0.141 & 0.347 &  & 0.041    & 0.097 & 0.245 \\
Oracle   & 0.184     & 0.868  & 1      &  & 0.133   & 0.631 & 0.991 &  & 0.123    & 0.517 & 0.976 \\
New & 0.157     & 0.666  & 0.971  &  & 0.112   & 0.394 & 0.810 &  & 0.112    & 0.340 & 0.686 \\ \hline
\end{tabular}%
\end{table}

\begin{table}[tbh!]
\tabcolsep 4pt
\centering
\caption{Empirical powers of Chen and Qin's test (CQ), Bai and Sarandasa's test (BS), Srivastava and Du's test (SD), the Oracle test, and the proposed Neighborhood-Assisted $T^2$ test (New) with randomly distributed $\mu$ and $\Sigma$ specified in model (c).}
\label{table6}
\begin{tabular}{cccclccclccc}
\hline
\multicolumn{12}{c}{$\beta=0.4$}                                                                  \\ \hline
         & \multicolumn{3}{c}{$p=200$} &  & \multicolumn{3}{c}{400} &  & \multicolumn{3}{c}{1000} \\ \cline{2-4} \cline{6-8} \cline{10-12}
         & $r=0.2$   & 0.4    & 0.6    &  & $r=0.2$ & 0.4   & 0.6   &  & $r=0.2$  & 0.4   & 0.6   \\ \hline
CQ       & 0.644     & 1      & 1      &  & 0.701   & 1     & 1     &  & 0.775    & 1     & 1     \\
BS       & 0.644     & 1      & 1      &  & 0.700   & 1     & 1     &  & 0.777    & 1     & 1     \\
SD       & 0.581     & 1      & 1      &  & 0.639   & 1     & 1     &  & 0.678    & 1     & 1     \\
Oracle   & 1         & 1      & 1      &  & 1       & 1     & 1     &  & 1        & 1     & 1     \\
New & 0.576     & 1      & 1      &  & 0.602   & 1     & 1     &  & 0.663    & 1     & 1     \\ \hline
\multicolumn{12}{c}{$\beta=0.8$}                                                                  \\ \hline
         & \multicolumn{3}{c}{$p=200$} &  & \multicolumn{3}{c}{400} &  & \multicolumn{3}{c}{1000} \\ \cline{2-4} \cline{6-8} \cline{10-12}
         & $r=0.2$   & 0.4    & 0.6    &  & $r=0.2$ & 0.4   & 0.6   &  & $r=0.2$  & 0.4   & 0.6   \\ \hline
CQ       & 0.095     & 0.274  & 0.750  &  & 0.084   & 0.197 & 0.491 &  & 0.074    & 0.151 & 0.380 \\
BS       & 0.097     & 0.274  & 0.748  &  & 0.083   & 0.194 & 0.493 &  & 0.074    & 0.152 & 0.380 \\
SD       & 0.079     & 0.236  & 0.680  &  & 0.060   & 0.157 & 0.439 &  & 0.038    & 0.090 & 0.277 \\
Oracle   & 0.236     & 0.998  & 1      &  & 0.214   & 0.795 & 1     &  & 0.137    & 0.791 & 1     \\
New & 0.087     & 0.232  & 0.522  &  & 0.071   & 0.172 & 0.319 &  & 0.072    & 0.121 & 0.234 \\ \hline
\end{tabular}%
\end{table}

\begin{table}[tbh!]
\tabcolsep 4pt
\centering
\caption{Empirical powers of Chen and Qin's test (CQ), Bai and Sarandasa's test (BS), Srivastava and Du's test (SD), the Oracle test, and the proposed Neighborhood-Assisted $T^2$ test (New) with randomly distributed $\mu$ and $\Sigma$ specified in model (d).}
\label{table7}
\begin{tabular}{cccclccclccc}
\hline
\multicolumn{12}{c}{$\beta=0.4$}                                                                  \\ \hline
         & \multicolumn{3}{c}{$p=200$} &  & \multicolumn{3}{c}{400} &  & \multicolumn{3}{c}{1000} \\ \cline{2-4} \cline{6-8} \cline{10-12}
         & $r=0.2$   & 0.4    & 0.6    &  & $r=0.2$ & 0.4   & 0.6   &  & $r=0.2$  & 0.4   & 0.6   \\ \hline
CQ       & 0.090     & 0.273  & 0.922  &  & 0.102   & 0.215 & 0.661 &  & 0.095    & 0.156 & 0.331 \\
BS       & 0.091     & 0.273  & 0.924  &  & 0.102   & 0.214 & 0.659 &  & 0.095    & 0.157 & 0.332 \\
SD       & 0.014     & 0.069  & 0.232  &  & 0.020   & 0.035 & 0.085 &  & 0.005    & 0.006 & 0.022 \\
Oracle   & 1         & 1      & 1      &  & 1       & 1     & 1     &  & 1        & 1     & 1     \\
New & 0.991     & 1      & 1      &  & 0.996   & 1     & 1     &  & 0.999    & 1     & 1     \\ \hline
\multicolumn{12}{c}{$\beta=0.8$}                                                                  \\ \hline
         & \multicolumn{3}{c}{$p=200$} &  & \multicolumn{3}{c}{400} &  & \multicolumn{3}{c}{1000} \\ \cline{2-4} \cline{6-8} \cline{10-12}
         & $r=0.2$   & 0.4    & 0.6    &  & $r=0.2$ & 0.4   & 0.6   &  & $r=0.2$  & 0.4   & 0.6   \\ \hline
CQ       & 0.063     & 0.084  & 0.109  &  & 0.062   & 0.084 & 0.089 &  & 0.068    & 0.083 & 0.086 \\
BS       & 0.063     & 0.084  & 0.110  &  & 0.063   & 0.084 & 0.089 &  & 0.068    & 0.083 & 0.086 \\
SD       & 0.015     & 0.019  & 0.024  &  & 0.007   & 0.015 & 0.011 &  & 0.006    & 0.007 & 0.007 \\
Oracle   & 0.232     & 0.917  & 1      &  & 0.175   & 0.782 & 1     &  & 0.142    & 0.667 & 1     \\
New & 0.162     & 0.651  & 0.841  &  & 0.117   & 0.415 & 0.764 &  & 0.107    & 0.322 & 0.631 \\ \hline
\end{tabular}%
\end{table}

Tables \ref{table4}-\ref{table7} demonstrate the empirical powers of all five tests with randomly distributed $\mu$  and subject to different structures of $\Sigma$ specified in models (a)-(d). For each case, the neighborhood size $k$ was chosen by the proposed stability neighborhood selection procedure. When signals were dense ($\beta=0.4$), all tests demonstrated good power performance under models (a)-(c). Since the key assumption about the covariance is not satisfied in model (d), the CQ, BS and SD tests had less power than the Neighborhood-Assisted $T^2$ and oracle tests. When signals were sparse ($\beta=0.8$), the powers of all tests had upward trend as both signal strength $r$ and dimension $p$ increased. Specially with models (a) and (b), the proposed Neighborhood-Assisted $T^2$ test had the power very close to that of the oracle test and much better than those of the CQ, BS and SD tests. This  confirms our theoretical findings in Theorem 3 that the proposed test is able to match the oracle test with $T_0^2(-1/2)$ with a bandable structure of $\Sigma$. With model (d), the Neighborhood-Assisted $T^2$ test was still able to incorporate partial dependence from the neighborhood such that its power was only slightly less than that of the oracle test. With model (c), the power of the proposed Neighborhood-Assisted $T^2$ test was less than that of the oracle test. But it was still comparable with the CQ, BS and SD tests. The reason the proposed test demonstrated less power is that the dependence described by the sparse covariance matrix in model (c) was not well exploited by the regularized estimator through banding the Cholesky factor. In Section 7, we will provide a solution to further improve the power of the Neighborhood-Assisted $T^2$ test under the sparse $\Sigma$.

Simulation results with clustering $\mu$ and non-Gaussian data were also conducted to demonstrate the non-parametric property of the proposed Neighborhood-Assisted $T^2$ test. Due to limited space, these results are included in the supplementary material of this paper.

\setcounter{section}{6} \setcounter{equation}{0}
\section*{\large 6. \bf Empirical Study}

Letrozole known as an anti-estrogen drug, is an aromatase inhibitor to treat postmenopausal breast cancer. To study its molecular effect on breast cancer, a microarray analysis was conducted to extract RNA from biopsies in 58 patients before and after 14-day treatment with letrozole (Miller et al., 2007). The normalized microarray data consist of pre-expression and post-expression of each of 22279 genes for each patient, and are available at http://www.ncbi.nlm.nih.gov/sites/GDSbrowser?acc=GDS3116. 

%The data have been analyzed by Miller et al.(2007) to provide insight into the molecular mechanism of action of letrozole, an anti-estrogen drug used to treat postmenopausal women with breast cancer. The original microarray gene expression data was generated through RNA extracted from paired tumour core biopsies taken before and after 14 days of treatment with letrozole in 58 patients (Miller WR, et al.,2007). Hence, we are interested in testing the hypothesis in (1.2), where ${\mu}$ represents the change of the gene expression level due to the treatment.

GO terms are gene-sets defined in Gene Ontology (GO) system that provides structured vocabularies to describe aspects of a gene product's biology. Our interest is to identify differentially expressed GO terms by testing the hypotheses (\ref{eq:hypo1}), where $\mu$ represents the difference of gene expression levels in each GO term before and after the treatment with letrozole. After the log-transformation of the original data, the GO terms were obtained by using the C5 collection of the GSEA online pathway databases (\url{http://software.broadinstitute.org/gsea/msigdb/collections.jsp#C5}). To accommodate high dimensionality, we further excluded some GO terms with the number of genes less than 60. As a result, there were 379 GO terms for our analysis.

%Table 6 lists the top ten GO terms that were identified by the proposed NA-$T^2$ test but not found by the CQ test. %These empirical results support our theoretically findings that the proposed combined test is more powerful that the CQ test by utilizing data dependence.

\begin{table}[t!]
\centering
\caption{The top 10 out of 24 GO terms that were declared significant by the proposed Neighborhood-Assisted $T^2$ test not by the CQ test.}
\label{table12}
\begin{tabular}{|c|c|c|}
\hline
GO ID   & GO Term Name                                                                                                         & P-Value      \\ \hline
0003713 & TRANSCRIPTION\_COACTIVATOR\_ACTIVITY                                                                                 & 8.068916e-16 \\ \hline
0008380 & RNA\_SPLICING                                                                                                        & 1.401153e-12 \\ \hline
0016071 & MRNA\_METABOLIC\_PROCESS                                                                                             & 8.155784e-12 \\ \hline
0032940 & SECRETION\_BY\_CELL                                                                                                  & 9.349589e-10 \\ \hline
0045045 & SECRETORY\_PATHWAY                                                                                                   & 1.605390e-09 \\ \hline
0005261 & CATION\_CHANNEL\_ACTIVITY                                                                                            & 8.814557e-08 \\ \hline
0007610 & BEHAVIOR                                                                                                             & 4.893933e-07 \\ \hline
0031966 & MITOCHONDRIAL\_MEMBRANE                                                                                              & 4.951206e-07 \\ \hline
0019933 & CAMP\_MEDIATED\_SIGNALING                                                                                            & 9.371950e-07 \\ \hline
0001775 & CELL\_ACTIVATION                                                                                                     & 1.132443e-06 \\ \hline
\end{tabular}
\end{table}

To meet the requirement that $k=o(n)$, we applied the stability selection procedure in Section 3.3 to select the optimal neighborhood size from $\{0, 1, \cdots, 8\}$. With the chosen neighborhood size for each of 379 GO terms, we further applied our proposed Neighborhood-Assisted $T^2$ test to obtain the corresponding P-value. To make a comparison, we also obtained the P-values from the CQ test. After controlling the family-wise error rate at 0.05 level with the Bonferroni correction, the CQ test declared 349 significant GO terms, which were all claimed to be significant by the proposed Neighborhood-Assisted $T^2$ test. %347 out of which were also declared significant by the proposed NA-$T^2$ test (the two exceptions are VOLTAGE\_GATED\_CHANNEL\_ACTIVITY and CYTOKINE\_ACTIVITY). 
On the other hand, the proposed Neighborhood-Assisted $T^2$ test found 24 more significant GO terms that were not identified by the CQ test. This is not surprising as the CQ test statistic can be thought as a special Neighborhood-Assisted $T^2$ test statistic with neighborhood size $k=0$. However, the Neighborhood-Assisted $T^2$ test can be more powerful than the CQ test by utilizing the neighborhood dependence. The top 10 GO terms that were identified by the proposed Neighborhood-Assisted $T^2$ test not by the CQ test were listed in Table 6.

\setcounter{section}{7} \setcounter{equation}{0}
\section*{\large 7. \bf Discussion}

In this article, we proposed a Neighborhood-Assisted $T^2$ test to revive the classical Hotelling's $T^2$ test in the ``large $p$, small $n$" paradigm. The Neighborhood-Assisted $T^2$ statistic was formulated by replacing $S_n^{-1}$ in the Hotelling's $T^2$ with the regularized covariance estimator through banding the Cholesky factor. The proposed test was able to mimic two oracle tests with known $\Sigma$ for the best power possible by adaptively adjusting its neighborhood size. A stability procedure was also provided to select the optimal neighborhood size via maximizing the signal-to-noise ratio. 

One may expect to revive the Hotelling's $T^2$ statistic by constructing other possible statistics obtained by replacing $S_n^{-1}$ with other high-dimensional estimators of $\Sigma^{-1}$ in the literature. For example, if the high-dimensional $\Sigma$ is  presumably sparse, the regularized estimator $\hat{\Sigma}^{-1}$ through banding or thresholding the sample covariance matrix, satisfies $||\hat{\Sigma}^{-1}-\Sigma^{-1}||=O_p\big\{({\log p}/{n})^{\alpha} \big\}$, where $||\cdot ||$ is the matrix $L_2$ (spectral) norm, and the constant $ \alpha \in (0, 1/2)$ 
(Bickel and Levina, 2008ab; Cai, Zhang and Zhou, 2010). A test statistic analogous to the proposed Neighborhood-Assisted $T^2$ is ${T}_m^{2} = n\bar{X}^{\prime}\hat{\Sigma}^{-1}\bar{X}$. To establish its asymptotic normality, we write
$$\frac{{T}_m^{2} - p}{\sqrt{2p}} = \frac{G_{n} - p}{\sqrt{2p}} + \frac{n\bar{X}^{\prime}(\hat{\Sigma}^{-1} -\Sigma^{-1})\bar{X}}{\sqrt{2p}},$$
where $G_{n} = n\bar{X}^{\prime}\Sigma^{-1}\bar{X}$. Under (\ref{model}), it can be shown that $(G_{n} - p)/{\sqrt{2p}} \xrightarrow{d} \mbox{N}(0, 1)$ as $p \to \infty$. However, unlike the Neighborhood-Assisted $T^2$ statistic that utilizes the regression analysis to control the error, the test statistic $T_m^2$ suffers from the error accumulation in $n\bar{X}^{\prime}(\hat{\Sigma}^{-1}-\Sigma^{-1})\bar{X}/{\sqrt{2p}}$. To see this, we notice that $n||\bar{X}||^2=O_p(p)$,  
\begin{eqnarray}
n\bar{X}^{\prime}(\hat{\Sigma}^{-1}-\Sigma^{-1})\bar{X}/{\sqrt{2p}}
&\le&n||\bar{X}||^2||\hat{\Sigma}^{-1}-\Sigma^{-1}||/{\sqrt{2p}}=O_p\biggl\{p^{\frac{1}{2}}(\frac{\log p}{n})^{\alpha} \biggr\},\nonumber
\end{eqnarray}
which is not negligible when $p > n^{2\alpha}$. Since $\alpha \in (0, 1/2)$, the asymptotic normality of $T_m^2$ is not established as $p>n$, and the revival of Hotelling's $T^2$ statistic based on $\hat{\Sigma}^{-1}$ ends in failure. 
 
The proposed Neighborhood-Assisted $T^2$ test can fully or partially exploit dependence to attain better power with bandable or equally correlated covariance matrix. However, the data dependence described by the sparse $\Sigma$ cannot be well exploited by $\hat{\Sigma}_k^{-1}$ in (\ref{e_matNonCenter}) as demonstrated in the simulation studies. One way to improve the power of the test with the sparse $\Sigma$ is to permute the $p$ components of the random vector $X=(X_1, \cdots, X_p)^{\prime}$ such that a bandable structure of the covariance of the permuted random vector appears. Note that the hypotheses (\ref{eq:hypo1}) do not change under such permutation. Specifically, given a sample of size $n$, one can randomly divide it into two subsamples with sizes $n_1$ and $n-n_1$, respectively. Using the subsample of size $n_1$, one can obtain a new ordering $\{1^*, \cdots, p^*\}$ such that the covariance of $X^*=(X_{1^*}, \cdots, X_{p^*})^{\prime}$ is approximately bandable. An Isomap algorithm can be found in Wagaman and Levina (2009) for this purpose. Based on the permuted observations in the second subsample by following the obtained ordering $\{1^*, \cdots, p^*\}$, we then construct the Neighborhood-Assisted $T^2$ statistic that is expected to achieve better performance in power. We will leave investigation about this to future study. 

It is worth mentioning the merit of $L_2$-norm test for its detectability against weak signals. Under the local alternative, by the signal to noise ratio (\ref{SNR_oracle}), the power of an $L_2$-norm test converges to 1 if $\mu' \mu \geq C_{0}p^{1/2} / n$ for a large constant $C_{0}$.
%Suppose all the signals have the common strength $\mu_{0}$ under 
Suppose that the number of non-zero entries of $\mu$ is $[p^{1-\beta}]$ where $[a]$ denotes the integer part of $a$ and $\beta \in (0, 1)$. Moreover, the non-zero entries are randomly distributed among $\{1, 2, \cdots, p\}$. The minimal signal strength that can be consistently detected by an $L_2$-norm test is at the order $p^{\beta/2 - 1/4}n^{-1/2}$.
When $\beta \in (0, 1/2)$ implying relatively dense signals, this level is at a smaller order of $n^{-1/2}$. 
Note that the minimum signal strength for higher criticism test and maximum test is at the order of $\sqrt{\log(p) / n}$. Thus, the $L_2$-norm test can be more powerful in the dense signal regime of $\beta \in (0, 1/2)$. %under (C2).

%\section*{ \bf Supplementary Material}

%The supplementary material provides the proofs of Lemmas, Proposition 1-4, and Theorems 1-3. It also includes more extensive simulation results to demonstrate the numerical performance of the proposed test. 

\setcounter{equation}{0}
\def\theequation{A.\arabic{equation}}
\def\thesection{A}

\section*{\large Appendix: Technical Details.}

%We begin with some lemmas that will be used to prove the main results. The technical proofs of these lemmas together with Propositions 1-4 are given in the supplementary material of this paper.

\noindent{\bf A.1. Proof of Theorem 1}

We first show that $\mbox{E}\{T_{N}^2(k)\}=p+n\mu^{\prime} \Sigma_k^{-1} \mu\{1+o(1)\}$. To prove it, we recall that $T_{N}^2(k)=\sum_{l=1}^p T_{N, l}^2(k)$, where, from the supplementary material,  
\bea
T_{N,l}^2 &=& \frac{ \{ \mathbf{1}^{\prime} (I - H_{l}) \mathscr{X}_{l} \}^{2} }{ {\mathscr{X}_{l}}^{\prime}(I - H_{l}) \mathscr{X}_{l} }. \nn
 \eea
With the condition that $\mu_i=o(k^{-1/2})$, it can be shown that $A_l=o(n)$ in Lemma 3 and $d_{k,l}=o(n)$ in Lemma 4. Then according to Lemmas 3 and 4 in supplementary material, we have
$\mbox{E}\{T_{N, l}^2(k)\}=1+n\alpha_l^2/d_{k,l}^2\{1+o(1)\}$.
As a result,  
\[
\mbox{E}\{T_{N}^2(k)\}=p+n\sum_{l=1}^p\alpha_l^2/d_{k,l}^2\{1+o(1)\}. 
\]
Using the definition of $\Sigma_k^{-1}$ in (\ref{Sigmak}) and $\alpha_l$ and $d_{k,l}^2$ right after (\ref{eq:RegInt}), we can derive that $\mbox{E}\{T_{N}^2(k)\}=p+n\mu^{\prime} \Sigma_k^{-1} \mu\{1+o(1)\}$. 

To prove Theorem 1, we define another statistic $T_U^2=n^{-1}(n-1)^{-1}\sum_{i \ne j}X_i^{\prime} \hat{\Sigma}_k^{-1}X_j\equiv \sum_{l=1}^p T_{U,l}^2$ where
\be
T_{U,l}^2 = \frac{1}{n(n-1)} \frac{ \sum_{i \neq j}^{n} e_{i}^{\prime} (I - H_{l}) \mathscr{X}_{l} {\mathscr{X}_{l}}^{\prime} (I - H_{l}) e_{j} }  { {\mathscr{X}_{l}}^{\prime}(I - H_{l}) \mathscr{X}_{l} / n }.\nonumber
\ee
It can be shown that $T_{U,l}^2=(T_{N, l}^2-1)/(n-1)$ or $T_{U}^2=(T_{N}^2-p)/(n-1)$. Therefore, to show Theorem 1, we only need to show that
\[
\frac{T_U^2-\mu^{\prime}\Sigma_k^{-1}\mu}{\sigma_U}\xrightarrow{d} \mbox{N}(0,1),
\]
where $\sigma_{U}^2=\frac{2}{n(n-1)} \mbox{tr}\{(\Sigma_{k}^{-1} \Sigma)^2\}+\frac{4}{n}\mu^{\prime}\Sigma_{k}^{-1} \Sigma \Sigma_{k}^{-1} \mu$.

To establish the normality of $T_U^2$, we consider another population version of $T_U^2$ 
\be
J_U^2=\frac{1}{n(n-1)}\sum_{i \ne j} X_i^{\prime} \Sigma_{k}^{-1} X_j, \label{CQ_t}
\ee
where $\Sigma_{k}^{-1}$ is defined in (\ref{Sigmak}). According to Chen and Qin (2010), it can be shown that under (\ref{model}) and condition (C1),
\[
\frac{J_U^2-\mu^{\prime} \Sigma_{k}^{-1}\mu}{\sigma_{U}}\xrightarrow{d} \mbox{N}(0,1) \quad \mbox{as}\quad p\to \infty \quad \mbox{and} \quad n \to \infty.
\]

Note that 
\[
\frac{T_U^2-\mu^{\prime} \Sigma_{k}^{-1}\mu}{\sigma_U}=\frac{J_U^2-\mu^{\prime} \Sigma_{k}^{-1}\mu}{\sigma_U}+\frac{T_U^2-J_U^2}{\sigma_U}.
\]
Therefore, to establish the normality of $T_U^2$, we only need to show that $\frac{T_U^2-J_U^2}{\sigma_U}=o_p(1)$.
Since ${\E} ( T_U^2)={\E}(J_U^2)$, we only need to show that $\mbox{Var}(T_U^2-J_U^2)=o(\sigma_U^2)$. 

Note that the test statistic $J_U^2$ can be written as 
\[
J_U^2=\sum_{l=1}^p J_{U,l}^2=\sum_{l=1}^p \frac{\sum_{i\ne j}^n (X_{il}-\sum_{q=l-k}^{l-1}A_{lq}^{\prime}X_{iq})(X_{jl}-\sum_{q=l-k}^{l-1}A_{lq}^{\prime}X_{jq})}{n(n-1)d_{k,l}^2}.
\]
Therefore, we only need to show that
\be
\sum_{l=1}^p \mbox{Var}(T_{U,l}^2-J_{U,l}^2)+ \sum_{l \ne q}^p \mbox{Cov}(T_{U,l}^2-J_{U,l}^2, T_{U,q}^2-J_{U,q}^2)= o(\sigma_U^2), \label{eq-normality}
\ee
which requires to show the following two results, separately:
\be
\sum_{l=1}^p \mbox{Var}(T_{U,l}^2-J_{U,l}^2)=o(\sigma_U^2),  \qquad \mbox{and} \label{eq-normality1}
\ee
\be
 \sum_{l \ne q}^p \mbox{Cov}(T_{U,l}^2-J_{U,l}^2, T_{U,q}^2-J_{U,q}^2)= o(\sigma_U^2). \label{eq-normality2}
\ee

Note that 
\begin{eqnarray}
& &\mbox{Var}(T_{U,l}^2-J_{U,l}^2) \ = \ \mbox{Var}(\frac{T_{N, l}^2}{n-1}-J_{U,l}^2)
 \ = \ \mbox{Var}\biggl[\frac{ \{ \mathbf{1}^{\prime} (I - H_{l}) \mathscr{X}_{l} \}^{2} }{(n-1) {\mathscr{X}_{l}}^{\prime}(I - H_{l}) \mathscr{X}_{l} } -J_{U,l}^2\biggr]\nonumber\\
&\le& 2\mbox{Var}\biggl[\frac{ \{ \mathbf{1}^{\prime} (I - H_{l}) \mathscr{X}_{l} \}^{2} }{(n-1) {\mathscr{X}_{l}}^{\prime}(I - H_{l}) \mathscr{X}_{l} } -\frac{ \{ \mathbf{1}^{\prime} (I - H_{l}) \mathscr{X}_{l} \}^{2} }{n(n-1) d_{k,l}^2 }\biggr]\nonumber\\
&+& 2\mbox{Var}\biggl[ \frac{ \{ \mathbf{1}^{\prime} (I - H_{l}) \mathscr{X}_{l} \}^{2} }{n(n-1) d_{k,l}^2 }- \frac{\sum_{i\ne j}^n (X_{il}-\sum_{q=l-k}^{l-1}A_{lq}^{\prime}X_{iq})(X_{jl}-\sum_{q=l-k}^{l-1}A_{lq}^{\prime}X_{jq})}{n(n-1)d_{k,l}^2}\biggr].\nonumber
\end{eqnarray}
Therefore, to show (\ref{eq-normality1}), we only need to show that  
\be
\mbox{Var}\biggl[\frac{ \{ \mathbf{1}^{\prime} (I - H_{l}) \mathscr{X}_{l} \}^{2} }{(n-1) {\mathscr{X}_{l}}^{\prime}(I - H_{l}) \mathscr{X}_{l} } -\frac{ \{ \mathbf{1}^{\prime} (I - H_{l}) \mathscr{X}_{l} \}^{2} }{n(n-1) d_{k,l}^2 }\biggr]=o(\sigma_U^2/p), \label{eq-normality3}
\ee
and
\be
\mbox{Var}\biggl[ \frac{ \{ \mathbf{1}^{\prime} (I - H_{l}) \mathscr{X}_{l} \}^{2} }{n(n-1) d_{k,l}^2 }- \frac{\sum_{i\ne j}^n (X_{il}-\sum_{q=l-k}^{l-1}A_{lq}^{\prime}X_{iq})(X_{jl}-\sum_{q=l-k}^{l-1}A_{lq}^{\prime}X_{jq})}{n(n-1)d_{k,l}^2}\biggr]=o(\sigma_U^2/p). \label{eq-normality4}
\ee

First, consider (\ref{eq-normality3}). By the fact that given $\mathcal{F}_l$, $\mathcal{G}_l$ follows an $F$-distribution with degrees of freedom $1$ and $n-k-1$ and non-central parameter $\mathcal{F}_l \alpha_l^2/ d_{k,l}^2$, we have
\begin{eqnarray}
\mbox{Var}\biggl[\frac{ \{ \mathbf{1}^{\prime} (I - H_{l}) \mathscr{X}_{l} \}^{2} }{(n-1) {\mathscr{X}_{l}}^{\prime}(I - H_{l}) \mathscr{X}_{l} } \biggr]%&=&\frac{1}{(n-1)^2}\mbox{Var}\biggl\{\frac{\mathcal{F}_l\mathcal{G}_l}{(n-k-1)+\mathcal{G}_l}\biggr\}\nonumber\\
%&=&\frac{1}{(n-1)^2}\biggl[{\E}\biggl\{\mathcal{F}_l^2\mbox{Var}\biggl(\frac{\mathcal{G}_l}{(n-k-1)+\mathcal{G}_l} | \mathcal{F}_l\biggr) \biggr\}  \nonumber\\
%&+& \mbox{Var}\biggl\{ \mathcal{F}_l \mbox{E}\biggl(\frac{\mathcal{G}_l}{(n-k-1)+\mathcal{G}_l} | \mathcal{F}_l\biggr)\biggr\}\biggr]\nonumber\\
&=&\frac{1}{(n-1)^2}\biggl\{2+4(n-k) \alpha_l^2/ d_{k,l}^2 \biggr\}\{1+o(1)\}. \label{eq-normality3-1}
\end{eqnarray}
Using the fact that given $\mathcal{F}_l$, $\frac{\{ \mathbf{1}^{\prime} (I - H_{l}) \mathscr{X}_{l} \}^{2}}{\mathcal{F}_l d_{k,l}^2} \sim \chi^2_1$, we have
\begin{eqnarray}
\mbox{Var}\biggl[ \frac{ \{ \mathbf{1}^{\prime} (I - H_{l}) \mathscr{X}_{l} \}^{2} }{n(n-1) d_{k,l}^2 }\biggr]%&=&\frac{1}{n^2(n-1)^2} \mbox{Var}\biggl(\mathcal{F}_l\frac{\{ \mathbf{1}^{\prime} (I - H_{l}) \mathscr{X}_{l} \}^{2}}{\mathcal{F}_l d_{k,l}^2} \biggr)\nonumber\\
%&=&\frac{1}{n^2(n-1)^2}\biggl[{\E}\biggl\{\mathcal{F}_l^2\mbox{Var}\biggl(\frac{\{ \mathbf{1}^{\prime} (I - H_{l}) \mathscr{X}_{l} \}^{2}}{\mathcal{F}_l d_{k,l}^2}| \mathcal{F}_l\biggr) \biggr\}  \label{eq-normality3-2}\\
%&+& \mbox{Var}\biggl\{ \mathcal{F}_l \mbox{E}\biggl(\frac{\{ \mathbf{1}^{\prime} (I - H_{l}) \mathscr{X}_{l} \}^{2}}{\mathcal{F}_l d_{k,l}^2} | \mathcal{F}_l\biggr)\biggr\}\biggr]\nonumber\\
&=&\frac{1}{n^2(n-1)^2}\biggl\{2(n-k)^2+4(n-k)^3 \alpha_l^2/ d_{k,l}^2 \biggr\}\{1+o(1)\}.\nonumber\\ 
\label{eq-normality3-2} 
\end{eqnarray}
And, using $\hat{\epsilon}_{l}^{\prime}\hat{\epsilon}_{l} = {\mathscr{X}_{l}}^{\prime}(I - H_{l})\mathscr{X}_{l} - \mathcal{F}_l^{-1} \{\mathbf{1}^{\prime}(I - H_{l})\mathscr{X}_{l}\}^{2}$ proved in the supplementary material, we have seen that 
\[
\frac{ \{ \mathbf{1}^{\prime} (I - H_{l}) \mathscr{X}_{l} \}^{2} }{{\mathscr{X}_{l}}^{\prime}(I - H_{l}) \mathscr{X}_{l} }=\mathcal{F}_l\biggl(\frac{\hat{\epsilon}_l^{\prime}\hat{\epsilon}_l/d_{k,l}^2}{\mathcal{F}_l \hat{\alpha}^2/d_{k,l}^2}+1 \biggr)^{-1}, 
\]
where given $\mathcal{F}_l$, $\hat{\epsilon}_l^{\prime}\hat{\epsilon}_l/d_{k,l}^2 \sim \chi^2_{n-k-1}$ is independent of $\mathcal{F}_l \hat{\alpha}_l^2/d_{k,l}^2 \sim \chi^2_1(\mathcal{F}_l \alpha_l^2/d_{k,l}^2)$. 
Similarly,
\[
\frac{ \{ \mathbf{1}^{\prime} (I - H_{l}) \mathscr{X}_{l} \}^{2} }{d_{k,l}^2 }=\mathcal{F}_l (\mathcal{F}_l \hat{\alpha}_l^2/d_{k,l}^2).
\]
If let $V_1=\mathcal{F}_l \hat{\alpha}_l^2/d_{k,l}^2$ and $V_2=\hat{\epsilon}_l^{\prime}\hat{\epsilon}_l/d_{k,l}^2$, then in (\ref{eq-normality3}), the covariance part is
\begin{eqnarray}
-\frac{2}{n(n-1)^2} \mbox{Cov}\biggl\{\mathcal{F}_l\biggl(\frac{V_2}{V_1}+1 \biggr)^{-1}, \mathcal{F}_l V_1 \biggr\}
&=&-\frac{2}{n(n-1)^2} \biggl[{\E}\biggl\{\mathcal{F}_l\mbox{Cov}(\frac{V_1}{V_1+V_2}, V_1|\mathcal{F}_l) \biggr\}\nonumber\\
&+&\mbox{Cov}\biggl\{ \mathcal{F}_l {\E}(\frac{V_1}{V_1+V_2}|\mathcal{F}_l), \mathcal{F}_l {\E}(V_1| \mathcal{F}_l)\biggr\} \biggr],\label{cov1}
\end{eqnarray}
where
\begin{eqnarray}
\mbox{Cov}(\frac{V_1}{V_1+V_2}, V_1|\mathcal{F}_l)&=& {\E}(\frac{V_1^2}{V_1+V_2}|\mathcal{F}_l)-{\E}(\frac{V_1}{V_1+V_2}|\mathcal{F}_l){\E}(V_1|\mathcal{F}_l).\label{cov2}
\end{eqnarray}
Since $V_1 \sim \chi^2_1(\lambda)$ with $\lambda=\mathcal{F}_l \alpha_l^2/d_{k,l}^2$ and $V_2 \sim \chi^2_{n-k-1}$, then
\begin{eqnarray}
{\E}(\frac{V_1^2}{V_1+V_2}|\mathcal{F}_l)%&=&\sum_{i=1}^{\infty} \frac{e^{-\lambda/2}(\lambda/2)^i}{i!} \int_0^{\infty} f(v_2)dv_2 \nonumber\\
%&\times& \int_{0}^{\infty}\frac{1}{\Gamma(\frac{1+2i}{2})2^{\frac{1+2i}{2}}} v_1^{\frac{1+2i}{2}-1}e^{-v_1/2}\frac{v_1^2}{v_1+v_2} dv_1\nonumber\\
&=&\sum_{i=1}^{\infty} \frac{e^{-\lambda/2}(\lambda/2)^i}{i!}(1+2i)\int_0^{\infty} f(v_2)dv_2\nonumber\\
&\times&\int_{0}^{\infty}\frac{1}{\Gamma(\frac{3+2i}{2})2^{\frac{3+2i}{2}}} v_1^{\frac{3+2i}{2}-1}e^{-v_1/2}\frac{v_1}{v_1+v_2} dv_1,\nonumber
\end{eqnarray} 
which, by the fact that $V_1/(V_1+V_2)$ follows Beta($\frac{3+2i}{2}, \frac{n-k-1}{2}$), is equal to
\[
\sum_{i=1}^{\infty} \frac{e^{-\lambda/2}(\lambda/2)^i}{i!} \frac{(1+2i)(3+2i)}{n-k+2+2i} \approx \frac{1}{n-k}(3+6\lambda+\lambda^2).
\]
Similarly, ${\E}(\frac{V_1}{V_1+V_2}|\mathcal{F}_l) \approx (n-k)^{-1}(1+\lambda)$ and ${\E}(V_1|\mathcal{F}_l)=1+\lambda$. As a result, (\ref{cov2}) becomes 
$(n-k)^{-1} (2+4\mathcal{F}_l \alpha_l^2/d_{k,l}^2)$, 
which is plugged into (\ref{cov1}) and leads to
\begin{eqnarray}
-\frac{2}{n(n-1)^2} \mbox{Cov}\biggl\{\mathcal{F}_l\biggl(\frac{V_2}{V_1}+1 \biggr)^{-1}, \mathcal{F}_l V_1 \biggr\}
%&=&-\frac{2}{n(n-1)^2(n-k)}\biggl\{ 2{\E}(\mathcal{F}_l^2)+4\alpha_l^2/d_{k,l}^2{\E}(\mathcal{F}_l^3)\nonumber\\
%&+&\mbox{Cov}(\mathcal{F}_l+\mathcal{F}_l^2 \alpha_l^2/d_{k,l}^2,\mathcal{F}_l+\mathcal{F}_l^2 \alpha_l^2/d_{k,l}^2)\biggr\}\nonumber\\
=-\frac{2}{n(n-1)^2} \{2(n-k)+4(n-k)^2\alpha_l^2/d_{k,l}^2 \}.\nonumber\\ \label{eq-normality3-3} 
\end{eqnarray}
Combining the results in (\ref{eq-normality3-1}), (\ref{eq-normality3-2}) and (\ref{eq-normality3-3}), we have 
\[
\mbox{Var}\biggl[\frac{ \{ \mathbf{1}^{\prime} (I - H_{l}) \mathscr{X}_{l} \}^{2} }{(n-1) {\mathscr{X}_{l}}^{\prime}(I - H_{l}) \mathscr{X}_{l} } -\frac{ \{ \mathbf{1}^{\prime} (I - H_{l}) \mathscr{X}_{l} \}^{2} }{n(n-1) d_{k,l}^2 }\biggr]=O(\frac{2k^2}{n^4}+\frac{4k^2}{n^3}\frac{\alpha_l^2}{d_{k,l}^2}).
\]
Therefore, (\ref{eq-normality3}) holds if $k=o(n)$.   

Next, we need to show that (\ref{eq-normality4}) holds if $k=o(n)$. By using the fact that $X_{il}-\sum_{q=l-k}^{l-1}A_{iq}^{\prime}X_{iq}=\alpha_l+\epsilon_{il}$, it is equivalent to showing that $n^{-4}d_{k,l}^{-4}[\mbox{Var}\{\{ \mathbf{1}^{\prime} (I - H_{l}) \mathscr{X}_{l} \}^{2}\}+\mbox{Var}\{\sum_{i\ne j}(\alpha_l+\epsilon_{il})(\alpha_l+\epsilon_{jl})\}-2\mbox{Cov}\{\{ \mathbf{1}^{\prime} (I - H_{l}) \mathscr{X}_{l} \}^{2}, \sum_{i\ne j}(\alpha_l+\epsilon_{il})(\alpha_l+\epsilon_{jl})\}]=o(\sigma_U^2/p)$. Toward this end, we first evaluate $\mbox{Var}\{\{ \mathbf{1}^{\prime} (I - H_{l}) \mathscr{X}_{l} \}^{2}\}$. Recall that $\mathbf{1}^{\prime} (I - H_{l}) \mathscr{X}_{l}=\mathcal{F}_l\hat{\alpha}_l$ given in the supplementary material, and moreover, given $\mathcal{F}_l$, $\sqrt{\mathcal{F}_l}(\hat{\alpha}_l-\alpha_l)/d_{k,l}\sim \mbox{N}(0, 1)$. Then, 
\begin{eqnarray}
\mbox{Var}\{\{ \mathbf{1}^{\prime} (I - H_{l}) \mathscr{X}_{l} \}^{2}\}%&=& \mbox{E}\{\mbox{Var}(\mathcal{F}_l^2 \hat{\alpha}_l^2|\mathcal{F}_l)\}+\mbox{Var}\{\mbox{E}(\mathcal{F}_l^2 \hat{\alpha}_l^2|\mathcal{F}_l) \}\nonumber\\
&=&\{4\alpha_l^2d_{k,l}^2(n-k)^3+2(n-k)^2d_{k,l}^4\}\{1+o(1)\}.\label{rr1}
\end{eqnarray}
And,
\begin{eqnarray}
\mbox{Var}\{\sum_{i\ne j}(\alpha_l+\epsilon_{il})(\alpha_l+\epsilon_{jl})\}%&=&\mbox{Var}(2n\alpha_l \sum_i \epsilon_{il}+\sum_{i \ne j} \epsilon_{il}\epsilon_{jl})\nonumber\\
%&=&4n^3\alpha_l^2\mbox{Var}(\epsilon_{il})+\mbox{E}\{(\sum_{i \ne j} \epsilon_{il}\epsilon_{jl})^2 \}\nonumber\\
&=&4n^3\alpha_l^2 d_{k,l}^2+2n(n-1)d_{k,l}^4.\label{rr2}
\end{eqnarray}
At last, using the fact that $(I - H_{l}) \mathscr{X}_{l}=\alpha_l(I - H_{l})\mathbf{1}+(I - H_{l})\epsilon_l$, where $\epsilon_l=(\epsilon_1, \cdots, \epsilon_p)^{\prime}$, we have
\begin{eqnarray}
&\quad&\mbox{Cov}\{\{ \mathbf{1}^{\prime} (I - H_{l}) \mathscr{X}_{l} \}^{2}, \sum_{i\ne j}(\alpha_l+\epsilon_{il})(\alpha_l+\epsilon_{jl})\}\nonumber\\
%&=&2(n-1)\alpha_l \biggl\{\mbox{E}(\mathcal{F}_l^2\hat{\alpha}_l^2\mathbf{1}^{\prime}\epsilon_l)+2\alpha_l \mbox{E}(\mathcal{F}_l \mathbf{1}^{\prime} (I - H_{l})\epsilon_l\mathbf{1}^{\prime}\epsilon_l)+\mbox{E}(\epsilon_l^{\prime}(I - H_{l})\mathbf{1}\mathbf{1}^{\prime}(I - H_{l})\epsilon_l\mathbf{1}^{\prime}\epsilon_l)\biggr\}\nonumber\\
%&+&\mbox{E}(\alpha_l^2\mathcal{F}_l^2\sum_{i \ne j}\epsilon_{il}\epsilon_{jl})+2\alpha_l \mbox{E}(\mathcal{F}_l\mathbf{1}^{\prime}(I - H_{l})\epsilon_l\sum_{i \ne j}\epsilon_{il}\epsilon_{jl})+\mbox{E}(\epsilon_l^{\prime}(I - H_{l})\mathbf{1}\mathbf{1}^{\prime}(I - H_{l})\epsilon_l\sum_{i \ne j}\epsilon_{il}\epsilon_{jl})\nonumber\\
&=&\{2(n-k)^2 d_{k,l}^4-2(n-k)d_{k,l}^4+4(n-1)(n-k)^2\alpha_l^2 d_{k,l}^2\}\{1+o(1)\}. \label{rr3}
\end{eqnarray}
Now combining (\ref{rr1}), (\ref{rr2}) and (\ref{rr3}), we see that
\begin{eqnarray}
&\quad &\mbox{Var}\biggl[ \frac{ \{ \mathbf{1}^{\prime} (I - H_{l}) \mathscr{X}_{l} \}^{2} }{n(n-1) d_{k,l}^2 }- \frac{\sum_{i\ne j}^n (X_{il}-\sum_{q=l-k}^{l-1}A_{lq}^{\prime}X_{iq})(X_{jl}-\sum_{q=l-k}^{l-1}A_{lq}^{\prime}X_{jq})}{n(n-1)d_{k,l}^2}\biggr]\nonumber\\
&=&O(\frac{2(1+2k)}{n^3}+\frac{4(k+2)}{n^2}\frac{\alpha_l^2}{d_{k,l}^2}). \nonumber
\end{eqnarray}
Therefore, (\ref{eq-normality4}) holds if $k=o(n)$. 

At last, we need to show that (\ref{eq-normality2}) is true. Toward this end, we only give a proof of (\ref{eq-normality2}) under the null hypothesis. The proof under the alternative with the requirement that $\mu_i=o(k^{-1/2})$ will be very similar. 

%Since $\sum_{l, q}^p \mbox{Cov}(J_{U,l}^2, J_{U,q}^2)=\sigma_U^2$ and $\mbox{Cov}(J_{U,l}^2, J_{U,q}^2) \ge 0$ for all $1\le l, q \le p$, we only need to show that
%\[
%\sum_{l\ne q}^p \mbox{Cov}(T_{U,l}^2-J_{U,l}^2, T_{U,q}^2-J_{U,q}^2)=o\{\sum_{l\ne q}^p\mbox{Cov}(J_{U,l}^2, J_{U,q}^2)\}.
%\]
%Moreover, due to the fact that $\mbox{E}(J_{U,l}^2)=0$ and $\mbox{E}(T_{U,l}^2)=0$, we only need to show that for each $l \ne q$, 
%\be
%\mbox{E}\biggl\{(T_{U,l}^2-J_{U,l}^2)(T_{U,q}^2-J_{U,q}^2)\biggr\}=o\biggl\{\mbox{E}(J_{U,l}^2 J_{U,q}^2)\biggr\}.\label{eq-normality5}
%\ee

First, we consider the case with neighborhood size $k=0$. With $k=0$,
\[
\frac{T_{N,l}^2}{n-1}-J_{U,l}^2=\frac{1}{n-1}+\frac{\sum_{i\ne j}X_{il}X_{jl}}{(n-1)\sum_i X_{il}^2}-\frac{\sum_{i\ne j}X_{il}X_{jl}}{n(n-1)\sigma_{ll}}.
\]
Using the result that $\mbox{E}(\sum_{i\ne j}X_{il}X_{jl}/\sum_i X_{il}^2)=0$ and $$\mbox{E}\biggl\{\frac{\sum_{i\ne j}X_{il}X_{jl}\sum_{i\ne j}X_{iq}X_{jq}}{\sum_i X_{il}^2\sum_i X_{iq}^2}\biggr\}=\frac{\mbox{E}(\sum_{i\ne j}X_{il}X_{jl}\sum_{i\ne j}X_{iq}X_{jq})}{\mbox{E}(\sum_i X_{il}^2\sum_i X_{iq}^2)}\{1+O(n^{-1})\},$$ we can show that 
$\mbox{E}\biggl\{(\frac{T_{N,l}^2}{n-1}-J_{U,l}^2)(\frac{T_{N,q}^2}{n-1}-J_{U,q}^2)\biggr\}%&=&\frac{1}{(n-1)^2}+\frac{2\sigma_{lq}^2}{n^2\sigma_{ll}\sigma_{qq}}\biggl\{ \frac{1}{1+2\sigma_{lq}^2/(n\sigma_{ll}\sigma_{qq})}-1\biggr\}\{1+o(1)\}\nonumber\\
={(n-1)^{-2}}-{4\sigma_{lq}^4}{n^{-3}\sigma_{ll}^{-2}\sigma_{qq}^{-2}}\{1+o(1)\}$. And $\mbox{E}(\frac{T_{N,l}^2}{n-1}-J_{U,l}^2)\mbox{E}(\frac{T_{N,q}^2}{n-1}-J_{U,q}^2)={(n-1)^{-2}}$. 
As a result, we have 
\[
\mbox{Cov}({T_{U,l}^2}-J_{U,l}^2, {T_{U,q}^2}-J_{U,q}^2)=-\frac{4\sigma_{lq}^4}{n^3\sigma_{ll}^2\sigma_{qq}^2}\{1+o(1)\},
\]
which is a smaller order of $\mbox{Cov}(J_{U,l}^2, J_{U,q}^2)=2n^{-1}\sigma_{lq}^2\sigma_{ll}^{-1}\sigma_{qq}^{-1}$. This implies that 
(\ref{eq-normality2}) holds for $k=0$. 

Next, we consider $k \ne 0$. Since $T_{U,l}^2=(T_{N,l}^2-1)/(n-1)$, to show (\ref{eq-normality2}), we need to show that 
\[
\sum_{l\ne q}^p\biggl[\mbox{E}\biggl\{(\frac{T_{N,l}^2}{n-1}-J_{U,l}^2)(\frac{T_{N,q}^2}{n-1}-J_{U,q}^2)\biggr\}-\frac{1}{(n-1)^2}\biggr]=o(\sigma_U^2).
\]

For this purpose, we need to evaluate $(n-1)^{-2}\mbox{E}(T_{N,l}^2 T_{N,q}^2)$, $(n-1)^{-1}\mbox{E}(T_{N,l}^2 J_{U,q}^2)$, $(n-1)^{-1}\mbox{E}(T_{N,q}^2 J_{U,l}^2)$, and $\mbox{E}(J_{U,l}^2 J_{U,q}^2)$, separately. First,
\begin{eqnarray}
\mbox{E}(J_{U,l}^2 J_{U,q}^2)=\frac{2}{n(n-1)d_{k,l}^2d_{k,q}^2}(\sigma_{lq}-\sum_{s}A_{qs}\sigma_{ls}-\sum_t A_{lt}\sigma_{qt}+\sum_{st}A_{lt}A_{qs}\sigma_{st})^2. \label{Tm1-1}
\end{eqnarray}

Next, we can show that 
\begin{eqnarray} 
\frac{\mbox{E}(T_{N,l}^2 J_{U,q}^2)}{n-1}%&=&\frac{1}{n(n-1)^2}\mbox{E}\biggl[\frac{\{\mathbf{1}^{\prime} (I - H_{l}) \mathscr{X}_{l}\}^2}{\mathscr{X}_{l}^{\prime} (I - H_{l}) \mathscr{X}_{l}} \frac{\sum_{i\ne j}^n (X_{iq}-\sum_{s}A_{qs}^{\prime}X_{is})(X_{jq}-\sum_{s} A_{qs}^{\prime}X_{js})}{d_{k,q}^2}\biggr]\nonumber\\
&=&\frac{1}{n(n-1)^2(n-k)d_{k,l}^2d_{k,q}^2}\biggl[2n(n-1)(\sigma_{lq}-\sum_s A_{qs} \sigma_{ls})^2\nonumber\\
&+&\frac{2(n-1)(n-2)(n-3)}{n}\{\sum_t A_{lt}(\sigma_{qt}-\sum_s A_{qs}\sigma_{st})\}^2-4(n-1)(n-2)\nonumber\\
&\times&(\sigma_{lq}-\sum_s A_{qs}\sigma_{ls})\{\sum_s A_{ls}(\sigma_{sq}-\sum_t A_{qt}\sigma_{st})\} \biggr]\{1+o(1)\}.\label{Tm1-2}
\end{eqnarray}
Similarly, we have
\begin{eqnarray} 
\frac{\mbox{E}(T_{N,q}^2 J_{U,l}^2)}{n-1}
&=&\frac{1}{n(n-1)^2(n-k)d_{k,l}^2d_{k,q}^2}\biggl[2n(n-1)(\sigma_{lq}-\sum_s A_{ls} \sigma_{qs})^2\nonumber\\
&+&\frac{2(n-1)(n-2)(n-3)}{n}\{\sum_t A_{qt}(\sigma_{lt}-\sum_s A_{ls}\sigma_{st})\}^2-4(n-1)(n-2)\nonumber\\
&\times&(\sigma_{lq}-\sum_s A_{ls}\sigma_{qs})\{\sum_s A_{qs}(\sigma_{sl}-\sum_t A_{lt}\sigma_{st})\} \biggr]\{1+o(1)\}.\label{Tm1-3}
\end{eqnarray}
 
At last, we have
\begin{eqnarray}
\frac{\mbox{E}(T_{N,l}^2 T_{N,q}^2)}{(n-1)^2}&=&\biggl\{\frac{1}{(n-1)^2}+\frac{2}{n(n-1)d_{k,l}^2d_{k,q}^2}(\sigma_{lq}-\sum_{s}A_{qs}\sigma_{ls}-\sum_t A_{lt}\sigma_{qt}\nonumber\\
&+&\sum_{st}A_{lt}A_{qs}\sigma_{st})^2 \biggr\}\{1+o(1)\}.\label{Tm1-4}
\end{eqnarray}
Combining (\ref{Tm1-1}), (\ref{Tm1-2}), (\ref{Tm1-3}) and (\ref{Tm1-4}), we can prove (\ref{eq-normality2}). This completes the proof of Theorem 1.

\noindent{\bf A.2. Proof of Theorem 2.}

\noindent To prove Theorem 2, we need to show that $\hat{\sigma}_{N,0}^2(k)/\mbox{tr}\{(\Sigma_{k}^{-1} \Sigma)^2\}=1+o_p(1)$. We only show it under the null hypothesis since the proof under the alternative hypothesis will be similar. Let's consider the first term in $\hat{\sigma}_{N,0}^2(k)$, which can be written as
\begin{eqnarray}
\frac{2\sum_{i\ne j}(X_i^{\prime}\hat{\Sigma}_k^{-1}X_j)^2}{n(n-1)}&=&\frac{2\sum_{i\ne j}(X_i^{\prime}{\Sigma}_k^{-1}X_j)^2}{n(n-1)}+\frac{4\sum_{i\ne j}X_i^{\prime}{\Sigma}_k^{-1}X_jX_i^{\prime}(\hat{\Sigma}_k^{-1}-\Sigma_k^{-1})X_j}{n(n-1)}\nonumber\\
&+&\frac{2\sum_{i\ne j}\{X_i^{\prime}(\hat{\Sigma}_k^{-1}-\Sigma_k^{-1})X_j\}^2}{n(n-1)}.\nonumber
\end{eqnarray}

Using the result in Chen and Qin (2010), we have 
\[
\frac{n^{-1}(n-1)^{-1}\sum_{i\ne j}(X_i^{\prime}{\Sigma}_k^{-1}X_j)^2}{\mbox{tr}\{(\Sigma_{k}^{-1} \Sigma)^2\}}=1+o_p(1).
\]
Therefore, we only need to show that 
\be
\frac{n^{-1}(n-1)^{-1}\sum_{i\ne j}\{X_i^{\prime}(\hat{\Sigma}_k^{-1}-\Sigma_k^{-1})X_j\}^2}{\mbox{tr}\{(\Sigma_{k}^{-1} \Sigma)^2\}}=o_p(1), \quad \mbox{and} \label{Tm2-1} 
\ee
\be
\frac{2n^{-1}(n-1)^{-1}\sum_{i\ne j}X_i^{\prime}{\Sigma}_k^{-1}X_jX_i^{\prime}(\hat{\Sigma}_k^{-1}-\Sigma_k^{-1})X_j}{\mbox{tr}\{(\Sigma_{k}^{-1} \Sigma)^2\}}=o_p(1). \label{Tm2-2}
\ee

For (\ref{Tm2-1}), we only need to show that
\[
\frac{\mbox{E}\{X_1^{\prime}(\hat{\Sigma}_k^{-1}-\Sigma_k^{-1})X_2\}^2}{\mbox{tr}\{(\Sigma_{k}^{-1} \Sigma)^2\}}=o(1).
\]
Toward this end, if $k=0$, $X_1^{\prime}(\hat{\Sigma}_k^{-1}-\Sigma_k^{-1})X_2=\sum_{l=1}^p nX_{1l}X_{2l}/\sum_{i=1}^n X_{il}^2-\sum_{l=1}^p X_{1l}X_{2l}/\sigma_{ll}$. Then, it can be shown that
$$\mbox{E}\{X_1^{\prime}(\hat{\Sigma}_k^{-1}-\Sigma_k^{-1})X_2\}^2=\sum_{l=1}^p O(n^{-1})+ \sum_{l\ne h} O\{n^{-1}\sigma_{lh}^4/(\sigma_{ll}^2\sigma_{hh}^2)\}=o[\mbox{tr}\{(\Sigma_{k}^{-1} \Sigma)^2\}].$$
If $k>0$, then, $\{X_1^{\prime}(\hat{\Sigma}_k^{-1}-\Sigma_k^{-1})X_2\}^2=(X_1^{\prime}\hat{\Sigma}_k^{-1}X_2)^2+(X_1^{\prime}{\Sigma}_k^{-1}X_2)^2-2X_1^{\prime}\hat{\Sigma}_k^{-1}X_2X_1^{\prime}{\Sigma}_k^{-1}X_2$, 
where, by letting $Q_l=\mathscr{X}_{l-k:l-1}^{\prime}\mathscr{X}_{l-k:l-1}$ and $A_l=\Sigma_{l-k:l-1}$,   
\begin{eqnarray}
X_1^{\prime}\hat{\Sigma}_k^{-1}X_2&=&\sum_{l=1}^p\biggl\{X_{1l}X_{2l}-\sum_{ab}Q^{-1}_{l, ab} (\sum_i X_{il} X_{ia})(X_{2b}X_{1l}+X_{1b}X_{2l})\nonumber\\
&+&\sum_{ab}\sum_{cd}Q^{-1}_{l, ab}Q^{-1}_{l, cd} (\sum_i X_{il} X_{ia})(\sum_i X_{il} X_{ic})X_{1b}X_{2d}\biggr\}\{\mathscr{X}_{l}^{\prime} (I - H_{l}) \mathscr{X}_{l}/n\}^{-1}, \nonumber
\end{eqnarray}
and
\[
X_1^{\prime}{\Sigma}_k^{-1}X_2=\sum_{l=1}^p \frac{X_{1l}X_{2l}+\sum_{ab}A^{-1}_{l, ab}\sigma_{la}(X_{2b}X_{1l}+X_{1b}X_{2l})+\sum_{ab}\sum_{cd}A^{-1}_{l, ab}A^{-1}_{l, cd}\sigma_{la}\sigma_{lc}X_{1b}X_{2d}}{d_{k,l}^2}.
\]

Using the above expression, we first have
\begin{eqnarray}
\mbox{E}\{(X_1^{\prime}{\Sigma}_k^{-1}X_2)^2\}%&=&\mbox{tr}\{(\Sigma_{k}^{-1} \Sigma)^2\}\nonumber\\
&=&\sum_{l,h}^p\frac{1}{d_{k,l}^2d_{k,h}^2}\biggl\{\sigma_{lh}^2+4\sum_{ab}A_{h,ab}^{-1}\sigma_{ha}\sigma_{lh}\sigma_{lb}+2\sum_{ab} \sum_{cd}A_{l,ab}^{-1}A_{l,cd}^{-1}\sigma_{la}\sigma_{lc}\sigma_{bh}\sigma_{dh}\nonumber\\
&+&2\sum_{ab} \sum_{cd}A_{l,ab}^{-1}A_{h,cd}^{-1}\sigma_{la}\sigma_{hc}(\sigma_{bd}\sigma_{lh}+\sigma_{ld}\sigma_{bh})\nonumber\\
&+&2\sum_{ab} \sum_{cd}\sum_{ef}A_{l,ab}^{-1}A_{l,cd}^{-1}A_{h,ef}^{-1}\sigma_{la}\sigma_{lc}\sigma_{he} (\sigma_{bh}\sigma_{df}+\sigma_{bf}\sigma_{dh})\nonumber\\
&+&\sum_{a_1b_1} \sum_{c_1d_1}\sum_{a_2b_2} \sum_{c_2d_2}A_{l,a_1b_1}^{-1}A_{l,c_1d_1}^{-1}A_{h,a_2b_2}^{-1}A_{h,c_2d_2}^{-1}\sigma_{la_1}\sigma_{lc_1}\sigma_{ha_2}\sigma_{hc_2}\sigma_{b_1b_2}\sigma_{d_1d_2}\biggr\}.\nonumber
\end{eqnarray}
Second, we have 
\begin{eqnarray}
\mbox{E}\{(X_1^{\prime}\hat{\Sigma}_k^{-1}X_2)^2\}&=&\sum_{l,h}^p\mbox{E}\biggl[\biggl\{X_{1l}X_{2l}-\sum_{ab}Q^{-1}_{l, ab} (\sum_i X_{il} X_{ia})(X_{2b}X_{1l}+X_{1b}X_{2l})\nonumber\\
&+&\sum_{ab}\sum_{cd}Q^{-1}_{l, ab}Q^{-1}_{l, cd} (\sum_i X_{il} X_{ia})(\sum_i X_{il} X_{ic})X_{1b}X_{2d}\biggr\}\nonumber\\
&\times&\biggl\{X_{1h}X_{2h}-\sum_{ab}Q^{-1}_{h, ab} (\sum_i X_{ih} X_{ia})(X_{2b}X_{1h}+X_{1b}X_{2h})\nonumber\\
&+&\sum_{ab}\sum_{cd}Q^{-1}_{h, ab}Q^{-1}_{h, cd} (\sum_i X_{ih} X_{ia})(\sum_i X_{ih} X_{ic})X_{1b}X_{2d}\biggr\}\biggr]\nonumber\\
&\times&\mbox{E}^{-1}[\{\mathscr{X}_{l}^{\prime} (I - H_{l}) \mathscr{X}_{l}\mathscr{X}_{h}^{\prime} (I - H_{h}) \mathscr{X}_{h}/n^2\}]\{1+o(1) \}, \nonumber
\end{eqnarray}
which is $\mbox{tr}\{(\Sigma_{k}^{-1} \Sigma)^2\}\{1+o(1)\}$. Similarly, we can show that 
\[
\mbox{E}(X_1^{\prime}\hat{\Sigma}_k^{-1}X_2X_1^{\prime}{\Sigma}_k^{-1}X_2)=\mbox{tr}\{(\Sigma_{k}^{-1} \Sigma)^2\}\{1+o(1)\}.
\]
As a result, we have 
$\mbox{E}\{X_1^{\prime}(\hat{\Sigma}_k^{-1}-\Sigma_k^{-1})X_2\}^2=o[\mbox{tr}\{(\Sigma_{k}^{-1} \Sigma)^2\}]$. 

Similarly, we can show that for $k=0$ and $k>0$,  
\[
\frac{\sum_{i_1\ne j_1}\sum_{i_2\ne j_2}\mbox{E}\{X_{i_1}^{\prime}{\Sigma}_k^{-1}X_{j_1}X_{i_1}^{\prime}(\hat{\Sigma}_k^{-1}-\Sigma_k^{-1})X_{j_1}X_{i_2}^{\prime}{\Sigma}_k^{-1}X_{j_2}X_{i_2}^{\prime}(\hat{\Sigma}_k^{-1}-\Sigma_k^{-1})X_{j_2} \}}{4n^{2}(n-1)^{2}\mbox{tr}^2\{(\Sigma_{k}^{-1} \Sigma)^2\}}=o(1),
\]
which implies that (\ref{Tm2-2}) is true. Similarly, we can show that the other two terms in $\hat{\sigma}_{N,0}^2(k)$ have the order $o_p[\mbox{tr}\{(\Sigma_{k}^{-1} \Sigma)^2\}]$. This completes the proof of Theorem 2.

\noindent{\bf A.3. Proof of Theorem 3}

The first claim is directly followed by (3.8). For the second claim,
note that the signal-to-noise ratio of the proposed test with  $k>0$ can be written as
\[
\mbox{SNR}_{N}(\mu, k)=\frac{n\mu^{\prime}\Sigma^{-1}\mu+\delta_1}{\sqrt{2p+4n\mu^{\prime}\Sigma^{-1}\mu+\delta_2+\delta_3}},
\]
where $\delta_1=n\mu^{\prime}(\Sigma_k^{-1}-\Sigma^{-1})\mu$, $\delta_2=4\mbox{tr}\{(\Sigma_k^{-1}-\Sigma^{-1})\Sigma\}+2\mbox{tr}\{(\Sigma_k^{-1}-\Sigma^{-1})\Sigma(\Sigma_k^{-1}-\Sigma^{-1})\Sigma\}$ and $\delta_3=8n\mu^{\prime}(\Sigma_k^{-1}-\Sigma^{-1})\mu+4n\mu^{\prime}(\Sigma_k^{-1}-\Sigma^{-1})\Sigma(\Sigma_k^{-1}-\Sigma^{-1})\mu$. So we only need to show that $\delta_1=o(n\mu^{\prime}\Sigma^{-1}\mu)$, $\delta_2=o(p)$ and $\delta_3=o(4n\mu^{\prime}\Sigma^{-1}\mu)$ as $k \to \infty$, respectively.   

For $\delta_1$, we notice that 
$|\delta_1| \le n ||\mu||^2||\Sigma_k^{-1}-\Sigma^{-1}||=O(n ||\mu||^2/k^{\alpha})$, 
where in the last step, we use the result $||\Sigma_k^{-1}-\Sigma^{-1}||=O(k^{-\alpha})$ from Bickel and Levina (2008). Therefore, we have $\delta_1=o(n\mu^{\prime}\Sigma^{-1}\mu)$ because all eigenvalues of $\Sigma$ are bounded when $\Sigma \in \mathcal{V}^{-1}$. Similarly, for $\delta_3$, we have
\[
|4n\mu^{\prime}(\Sigma_k^{-1}-\Sigma^{-1})\Sigma(\Sigma_k^{-1}-\Sigma^{-1})\mu| \le 4n ||\mu||^2 ||\Sigma_k^{-1}-\Sigma^{-1}||^2 ||\Sigma||=o(4n\mu^{\prime}\Sigma^{-1}\mu).
\]
%Thus, we have $\delta_3=o(4n\mu^{\prime}\Sigma^{-1}\mu)$. 
Next, we show that $\delta_2=o(p)$. By letting $A=\Sigma_k^{-1}-\Sigma^{-1}$, it can be shown that
\[
|\mbox{tr}\{(\Sigma_k^{-1}-\Sigma^{-1})\Sigma\}| \le \sum_{i}^p\sum_j^p|A_{ij}||\sigma_{ij}| \le p \max_{ij}|\sigma_{ij}| \max_i \sum_j |A_{ij}|=\max_{ij}|\sigma_{ij}| O(p/k^{\alpha}), 
\]
which is at an order of $o(p)$. Similarly, using $\mbox{tr}\{(\Sigma_k^{-1}-\Sigma^{-1})\Sigma(\Sigma_k^{-1}-\Sigma^{-1})\Sigma\} \le \mbox{tr}\{(\Sigma_k^{-1}-\Sigma^{-1})^2\Sigma^2 \}$, we can show that it is a smaller order of $p$. Thus, we have $\delta_2=o(p)$. This completes the proof of Theorem 3.

\section*{Reference}
\begin{description}
\item
{Anderson, T. W.} (2003), \textit{An Introduction to Multivariate Statistical Analysis}, Hoboken, NJ: Wiley.

%\textsc{Ashburner, M., Ball, C.A., Blake, J.A., Botstein, D., Butler, H., Cherry, J.M., Davis, A., Dolinski, K., Dwight, S.S., Eppig, J.T., et al.} (2000), ``The Gene Ontology Consortium. Gene Ontology: tool for the unification of biology'', \textit{ Nature Genetics}, 25, 25-29.

%\textsc{Bai, Z.} (1993), ``Convergence Rate of Expected Spectral Distributions of Large Random Matrices. Part II. Sample Covariance Matrices'', \textit{The Annals of Probability}, 21, 649-672.

%\textsc{Bai, Z., Yin, Y. Q.} (1993), ``Limit of the Smallest Eigenvalue of Large Dimensional Covariance Matrix'', \textit{The Annals of Probability}, 21, 1275-1294.

%\textsc{Bai, Z., Jiang, D., Yao, J. and Zheng, S.} (2009), ``Corrections to LRT on Large-Dimensional Covariance Matrix by RMT'', \textit{The Annals of Statistics}, 37, 3822-3840.
\item
{Bai, Z., and Saranadasa, H.} (1996),
``Effect of High Dimension: By an Example
of a Two Sample Problem'', \textit{Statistica Sinica}, 6, 311-329.

%\textsc{Bai, Z. and Silverstein, J.} (2010), \textit{Spectral Analysis of Large Dimensional Random Matrices}, Springer.
\item
{Barry, W., Nobel, A., and Wright, F.} (2005), ``Significance Analysis of Functional Categories in Gene Expression Studies: A Structured Permutation Approach'', \textit{Bioinformatics}, 21, 1943-1949.

%\textsc{Benjamini, Y., and Hochberg, Y.} (1995), ``Controlling the False Discovery Rate: A Practical and Powerful Approach to Multiple Testing'', \textit{Journal of the Royal Statistical Society Series B}, 57, 289-300.
\item
{Bickel, P. J. and Levina, E.} (2008a), ``Regularized Estimation of Large Covariance Matrices'', \textit{The Annals of Statistics}, 36, 199-227.
\item
{Bickel, P. J. and Levina, E.} (2008b), ``Covariance Regularization by Thresholding'', \textit{The Annals of Statistics}, 36, 2577-2604.

%\textsc{Cai, T., Liu, W. and Luo, X.} (2011), ``A Constrained $l_1$ Minimization Approach to Sparse Precision Matrix Estimation'', \textit{Journal of the American Statistical Association}, 106, 594-607.

\item
{Cai, T., Liu, W. and Xia, Y.} (2014), ``Two-Sample Test of High Dimensional Means under Dependence'', \textit{Journal of the Royal Statistical Society: Series B}, 76, 349-372.
\item
{Cai, T., Zhang, C. and Zhou, H.} (2010), ``Optimal Rates of Convergence for Covariance Matrix Estimation'', \textit{The Annals of Statistics}, 38, 2118-2144.

\item
{Chen, S. X., Li, J. and Zhong, P.} (2015),
`` Two-Sample and ANOVA Tests for High Dimensional Means'', Manuscript.
\item
{Chen, L., Paul, D., Prentice R. and Wang, P.} (2011),
``A Regularized Hotelling's T2 Test for Pathway Analysis in Proteomic Studies'', \textit{Journal of the American Statistical Association}, 106, 1345-1360.
\item
{Chen, S. X., and Qin, Y.-L.} (2010),
``A Two Sample Test for High Dimensional
Data With Applications to Gene-Set Testing'', \textit{The Annals of Statistics}, 38,
808-835.

%\textsc{Chen, S. X., Zhang, L.-X., and Zhong, P.-S.} (2010), ``Testing for High Dimensional Covariance Matrices'', \textit{Journal of the American Statistical Association}, 109, 810-819.

%\textsc{Chiaretti, S., Li, X. C., Gentleman, R., Vitale, A., Vignetti, M., Mandelli, F., Ritz, J., and Foa, R.} (2004), ``Gene Expression Profile of Adult T-Cell Acute Lymphocytic Leukemia Identifies Distinct Subsets of Patients with Different Response to Therapy and Survival '', \textit{Blood}, 103, 2771-2778.

%\textsc{Donoho, D., and Jin, J.} (2004), ``Higher Criticism for Detecting Sparse Heterogeneous Mixtures'', \textit{The Annals of Statistics}, 32, 962-994.

%\textsc{Dudoit, S., Keles, S., and van der Laan, M.} (2008), ``Multiple Tests of Association with Biological Annotation Metadata'', \textit{Institute of Mathematical Statistics Collections}, 2, 153-218.

%\textsc{Dykstra, R.L.} (1970), ``Establishing the Positive Definiteness of the Sample Covariance Matrix'', \textit{The Annals of Mathematical Statistics}, 41, 2153-2154.
\item
{Efron, B., and Tibshirani, R.} (2007), ``On Testing the Significance of Sets of Genes'',
\textit{The Annals of Applied Statistics}, 1, 107-129.

%\textsc{El Karoui, N.} (2007) ``Tracy-Widom Limit for the Largest Eigenvalue of a Large Class of Complex Sample Covariance Matrices'', \textit{The Annals of Probability}, 35, 663-714.

%\textsc{Fan, J., Fan, Y. and Lv, J.} (2008), ``High Dimensional Covariance Matrix Estimation Using a Factor model'', \textit{Journal of Econometrics}, 147, 186-197.

%\textsc{Fan, J., Hall, P., and Yao, Q.} (2007), ``How Many Simultaneous Hypothesis Tests Can Normal, Student's t or Bootstrap Calibration Be Applied'', \textit{Journal of the American Statistical Association}, 102, 1282-1288.

%\textsc{Fan, J., Peng, H., and Huang, T.} (2005), ``Semilinear High-Dimensional Model for Normalization of Microarray Data: A Theoretical Analysis and Partial Consistency'', \textit{Journal of the American Statistical Association}, 100, 781-796.

%\textsc{Feng, L., Zou, C., Wang, Z. and Zhu, L.} (2014), ``Two-sample Behrens-Fisher Problem for High-Dimensional Data'', \textit{Statistica Sinica}, preprint.

%\textsc{Hall, P., and Jin, J.} (2008), ``Properties of Higher Criticism Under Long-Range Dependence'', \textit{The Annals of Statistics}, 36, 381-402.
\item
{Hall, P., and Jin, J.} (2010), ``Innovated Higher Criticism for Detecting Sparse Signals in Correlated Noise'', \textit{The Annals of Statistics}, 38, 1686-1732.

%\textsc{Huang, J., Liu, N., Pourahmadi, M., and Liu, L.} (2006), ``Covariance Matrix Selection and Estimation via Penalised Normal Likelihood'', \textit{Biometrika}, 93, 85-98.

%\textsc{Huang, J., Wang, D., and Zhang, C.} (2005), ``A Two-way Semilinear Model for Normalization and Analysis of cDNA Microarray Data'', \textit{Journal of the American Statistical Association}, 100, 814-829.

%\textsc{Johnstone, I. M.} (2001), ``On the Distribution of the Largest Eigenvalue in Principal Components Analysis'', \textit{The Annals of Statistics}, 29, 295-327.

%\textsc{Johnstone, I. M., and Lu, A.} (2009), ``On Consistency and Sparsity for Principal Components Analysis in High Dimensions'', \textit{Journal of the American Statistical Association}, 104, 682-693.

%\textsc{Lam, C., and Yao, Q.} (2011), ``Factor Modelling for High-Dimensional Time Series: A Dimension-Reduction Approach'', Technical Report.

%\textsc{Lam, C., Yao, Q., and Bathia, N.} (2011), ``Estimation of Latent Factors for High-Dimensional Time Series'', Technical Report.

%\textsc{Lan, W., Luo, R., Tsai, C., Wang, H., and Yang, Y.} (2010), ``Testing the Diagonality of a Large Covariance Matrix in a Regression Setting'', Technical Report.

%\textsc{Ledoit, O., and Wolf, M.} (2002), ``Some Hypothesis Tests for the Covariance Matrix When the Dimension Is Large Compare to the Sample Size'', \textit{The Annals of Statistics}, 30, 1081-1102.

%\textsc{Ledoit, O., and Wolf, M.} (2004), ``A Well Conditioned Estimator for Large-Dimensional Covariance Matrices'', \textit{The Journal of Multivariate Analysis}, 88, 365-411.
\item
{Li, H., Aue, A., Paul D., Peng, J. and Wang, P.} (2016), ``An Adaptable Generalization of Hotelling's T2 Test in High Dimension'', Manuscript.
\item
{Li, J., and Chen, S. X.} (2012), ``Two Sample Tests for High Dimensional Covariance Matrices'', \textit{The Annals of Statistics}, 40, 908-940.
\item
{Miller, W., Larionov, A., Renshaw, L., Anderson, T., White, S., Murray, J., Murray, E., Hampton, G., Walker, J., Ho, S., Krause, A., Evans, D. and Dixon, J.} (2007), ``Changes in Breast Cancer Transcriptional Profiles after Treatment with the Aromatase Inhibitor, Letrozole'', \textit{Pharmacogenetics and Genomics}, 17, 813-826.

%\textsc{Nettleton, D., Recknor, J., and Reecy, J.} (2008), ``Identification of Differentially
%Expressed Gene Categories in Microarray Studies Using Nonparametric Multivariate
%Analysis'', \textit{Bioinformatics}, 24, 192-201.
\item
{Newton, M., Quintana, F., Den Boon, J., Sengupta, S., and Ahlquist, P.}
(2007), ``Random-Set Methods Identify Distinct Aspects of the Enrichment Signal in
Gene-Set Analysis'', \textit{The Annals of Applied Statistics}, 1, 85-106.

%\textsc{Rothman, A., Levina, L., and Zhu, J.} (2010), ``A New Approach to Cholesky-based Covariance Regularization in High Dimensions'', \textit{Biometrika}, 97, 539-550.

%\textsc{Schott, J.R.} (2007), ``Some High Dimensional Tests for a One-Way MANOVA'', \textit{The Journal of Multivariate Analysis}, 98, 1825-1839.

%\textsc{Schott, J.R.} (2007), ``A Test for the Equality of Covariance Matrices When the Dimension is Large Relative to the Sample Sizes'', \textit{Computational Statistics \& Data Analysis}, 51, 6535-6542.

%\textsc{Serfling, R. J.} (1980), \textit{Approximation Theorems of Mathematical Statistics}, John Wiley and Sons.

%\textsc{Shedden, K. and Taylor, J.} (2004), ``Differential Correlation Detects Complex Associations Between Gene Expression and Clinical Outcomes in Lung Adenocarcinomas'', \textit{Methods of Microarray Data Analysis IV}, Springer.
\item
{Srivastava, M.S. and Du, M.} (2008), ``A Test for the Mean Vector with Fewer Observations than the Dimension", \textit{Journal of Multivariate Analysis}, 99, 386-402.
\item
{Srivastava, R., Li, P. and Ruppert, D.} (2016), ``RAPTT: An Exact Two-Sample Test in High Dimensions Using Random Projections", \textit{Journal of Computational and Graphical Statistics}, 25, 954-970.

\item
{Thulin, M.} (2014), ``A High-Dimensional Two-Sample Test for the Mean Using Random Subspaces'', \textit{Computational Statistics and Data Analysis}, 74, 26-38.

%\textsc{van der Laan, M., and Bryan, J.} (2001), ``Gene Expression Analysis With the Parametric Bootstrap'', \textit{Biostatistics}, 2, 445-461.
\item
{Wagaman, A. S. and Levina, E.} (2009), ``Discovering Sparse Covariance Structures with the Isomap", \textit{Journal of Computational and Graphical Statistics}, 18, 551-572.
\item
{Wang, L., Peng, B. and Li, R.} (2015), ``A High-Dimensional Nonparametric Multivariate Test for Mean Vector", \textit{Journal of the American Statistical Association}, 110, 1658-1669. 
\item
{Zhong, P., Chen, S. X. and Xu, M.} (2013), ``Tests Alternative to Higher Criticism for High Dimensional Means under Sparsity and Column-wise Dependence'', \textit{The Annals of Statistics}, 41, 2820-2851.

\end{description}
\end{document}